\documentclass[journal]{IEEEtran}

\usepackage{lipsum}
\usepackage[pdftex]{graphicx}
\usepackage{bm,amsmath,amssymb,mathrsfs,amsfonts}
\usepackage{caption}
\usepackage{scalefnt,multirow}
\usepackage{here}
\usepackage{booktabs,arydshln}
\usepackage{enumitem}
\usepackage{type1cm}
\usepackage{cite}
\usepackage{url}
\usepackage{algorithm,algpseudocode}
\usepackage{color}
\usepackage{etoolbox}

\newcommand{\colorblue}{black}

\newcommand{\ie}{\textit{i.e.}}
\newcommand{\eg}{\textit{e.g.}}
\newcommand{\etal}{\textit{et~al.}}
\newcommand{\argmax}{\mathop{\rm argmax}\limits}
\newcommand{\exponential}{{\rm exp}}
\newcommand{\speech}{\bm{x}}
\newcommand{\encout}{\bm{h}}
\newcommand{\transcript}{\bm{y}}
\newcommand{\xmax}{T}
\newcommand{\hmax}{T'}
\newcommand{\ymax}{U}
\newcommand{\vocab}{{\cal V}}
\newcommand{\alignment}{\bm{A}}
\newcommand{\blocksizecurrent}{N_{\rm c}}
\newcommand{\blocksizeright}{N_{\rm r}}
\newcommand{\chunkmocha}{w}
\newcommand{\alphaij}{\alpha_{i,j}}
\newcommand{\alphadecotij}{\alpha^{\rm decot}_{i,j}}
\newcommand{\betaij}{\beta_{i,j}}
\newcommand{\pij}{p_{i,j}}
\newcommand{\eij}{e_{i,j}}
\newcommand{\uij}{u_{i,j}}
\newcommand{\zij}{z_{i,j}}
\newcommand{\cumsum}{{\tt cumsum}}
\newcommand{\cumprod}{{\tt cumprod}}
\newcommand{\movsum}{{\tt MovSum}}
\newcommand{\bmochai}{{\rm b}^{\rm mocha}_{i}}
\newcommand{\bctci}{{\rm b}^{\rm ctc}_{i}}
\newcommand{\bctc}{{\mathbf b}^{\rm ctc}}
\newcommand{\drefi}{{\rm b}^{\rm ref}_{i}}
\newcommand{\bref}{{\mathbf b}^{\rm ref}}
\newcommand{\blank}{\phi}
\newcommand{\ctcpath}{\bm{\pi}}
\newcommand{\probctc}{P_{\rm ctc}}
\newcommand{\probmocha}{P_{\rm mocha}}
\newcommand{\lambdaqua}{\lambda_{\rm qua}}
\newcommand{\lambdaminlt}{\lambda_{\rm minlt}}
\newcommand{\lambdactc}{\lambda_{\rm ctc}}
\newcommand{\lambdasync}{\lambda_{\rm sync}}
\newcommand{\deltadecot}{\delta_{\rm decot}}
\newcommand{\losstotal}{{\cal L}_{\rm total}}
\newcommand{\lossmocha}{{\cal L}_{\rm mocha}}
\newcommand{\lossqua}{{\cal L}_{\rm qua}}
\newcommand{\lossquadecot}{{\cal L}^{\rm decot}_{\rm qua}}
\newcommand{\lossminlt}{{\cal L}_{\rm minlt}}
\newcommand{\lossctc}{{\cal L}_{\rm ctc}}
\newcommand{\losssync}{{\cal L}_{\rm sync}}
\newcommand{\tcat}{T_{\rm cat}}
\newcommand{\specaugfreq}{F_{\rm sp}}
\newcommand{\specaugtime}{T_{\rm sp}}

\usepackage{pifont}%
\newcommand{\cmark}{\ding{51}}%
\newcommand{\xmark}{\ding{55}}%

\makeatletter
\def\adl@drawiv#1#2#3{%
        \hskip.5\tabcolsep
        \xleaders#3{#2.5\@tempdimb #1{1}#2.5\@tempdimb}%
                #2\z@ plus1fil minus1fil\relax
        \hskip.5\tabcolsep}
\newcommand{\cdashlinelr}[1]{%
  \noalign{\vskip\aboverulesep
           \global\let\@dashdrawstore\adl@draw
           \global\let\adl@draw\adl@drawiv}
  \cdashline{#1}
  \noalign{\global\let\adl@draw\@dashdrawstore
           \vskip\belowrulesep}}
\makeatother

\begin{document}
\title{Alignment Knowledge Distillation for Online Streaming Attention-based Speech Recognition}

\author{Hirofumi~Inaguma,~\IEEEmembership{Student Member,~IEEE,}
        and~Tatsuya~Kawahara,~\IEEEmembership{Fellow,~IEEE}%
\thanks{H. Inaguma and T. Kawahara are with the
Graduate School of Informatics, Kyoto University, Kyoto 606-8501, Japan (e-mail: \{inaguma,kawahara\}@sap.ist.i.kyoto-u.ac.jp).}%
}

\markboth{}%
{Shell \MakeLowercase{\etal}: Bare Demo of IEEEtran.cls for IEEE Journals}

\maketitle
\begin{abstract}
This article describes an efficient training method for online streaming attention-based encoder-decoder (AED) automatic speech recognition (ASR) systems.
AED models have achieved competitive performance in offline scenarios by jointly optimizing all components.
They have recently been extended to an online streaming framework via models such as monotonic chunkwise attention (MoChA).
However, the elaborate attention calculation process is not robust for long-form speech utterances.
Moreover, the sequence-level training objective and time-restricted streaming encoder cause a nonnegligible delay in token emission during inference.
To address these problems, we propose \textit{CTC synchronous training (CTC-ST)}, in which CTC alignments are leveraged as a reference for token boundaries to enable a MoChA model to learn optimal monotonic input-output alignments.
We formulate a purely end-to-end training objective to synchronize the boundaries of MoChA to those of CTC.
The CTC model shares an encoder with the MoChA model to enhance the encoder representation.
Moreover, the proposed method provides alignment information learned in the CTC branch to the attention-based decoder.
Therefore, CTC-ST can be regarded as \textit{self-distillation} of alignment knowledge from CTC to MoChA.
Experimental evaluations on a variety of benchmark datasets show that the proposed method significantly reduces recognition errors and emission latency simultaneously.
The robustness to long-form and noisy speech is also demonstrated.
We compare CTC-ST with several methods that distill alignment knowledge from a hybrid ASR system and show that the CTC-ST can achieve a comparable tradeoff of accuracy and latency without relying on external alignment information.
The best MoChA system shows recognition accuracy comparable to that of RNN-transducer (RNN-T) while achieving lower emission latency.
\end{abstract}


\IEEEpeerreviewmaketitle

\section{Introduction}
\IEEEPARstart{O}{nline} streaming automatic speech recognition (ASR) is a core technology for speech applications such as live captioning, simultaneous translation, voice search, and dialogue systems.
The traditional but still dominant approach in production is a hybrid system that modularizes the entire system into an acoustic model, a pronunciation model, and a language model (LM).
Recently, end-to-end (E2E) systems have achieved comparable performance to that of hybrid systems by optimizing a direct mapping function from the input speech to the target transcription~\cite{google_sota_asr,karita2019comparative,sainath2020streaming}.
Representative approaches include the connectionist temporal classification (CTC)~\cite{ctc_graves}, recurrent neural network transducer (RNN-T)~\cite{rnn_transducer}, recurrent neural aligner (RNA)~\cite{recurrent_neural_aligner}, hybrid autoregressive transducer (HAT)~\cite{variani2020hybrid}, and attention-based encoder-decoder (AED)~\cite{chorowski2015attention,chan2016listen} models.
With the simplified architecture, the E2E approaches are advantageous for rapid system development, on-device applications with a small footprint, and fast inference when a large amount of training data is available.

The E2E models have been compared in offline scenarios~\cite{s2s_comparison_google,s2s_comparison_baidu,huang2019exploring}, and AED models are typically the best choice because of the strong token dependency on the decoder side.
In online streaming scenarios, however, AED models are not suitable because they require the entire input in order to generate the initial token.
On the other hand, \textit{frame-synchronous} models such as CTC and RNN-T can easily be extended to the streaming setting.
RNN-T has been a practical choice because of its better performance than CTC with the help of token dependency modeling in the prediction network~\cite{he2019streaming,sainath2020streaming,li2020comparison}.
However, it is known that RNN-T consumes significant memory during training~\cite{bagby2018efficient,li2019improving} and requires a large search space during inference because of its frame-wise prediction, which significantly slows down the decoding speed.

To make AED models streamable, various methods have been proposed: neural transducer (NT)~\cite{neural_transducer}, triggered attention~\cite{moritz2019triggered_icassp2019}, adaptive computation steps (ACS)~\cite{adaptive_computation_steps}, continuous integrate-and-fire (CIF)~\cite{cif}, streaming chunk-aware multi-head attention (SCAMA)~\cite{zhang2020}, local windowing~\cite{luong2015effective,chan2016online,tjandra2017local,hou2017gaussian}, Gaussian mixture model (GMM) attention~\cite{graves2013generating}, neural autoregressive transducer (NAT)~\cite{luo2017learning}, and monotonic chunkwise attention (MoChA)~\cite{mocha}.
The major difference among these variants is the location where the input speech is segmented: at the encoder~\cite{neural_transducer,moritz2019triggered_icassp2019,adaptive_computation_steps,cif,zhang2020} or the decoder~\cite{luong2015effective,chan2016online,tjandra2017local,hou2017gaussian,graves2013generating,luo2017learning,mocha}).
Among these methods, we focus on MoChA because it can generate tokens with linear-time complexity at test time and can be trained as efficiently as offline AED models~\cite{kim2020attention,online_hybrid_ctc_attention,adaptive_mocha,inaguma2020streaming,online_hybrid_ctc_attention_taslp2020}.
Moreover, it can detect token boundaries by using contextual information captured in the decoder.

The main problem in making streaming AED models for practical systems is that they are not robust for long-form speech, which is not as problematic in frame-synchronous models~\cite{chiu2019comparison}.
Moreover, latency in the decision boundaries on token emission occurs in any E2E model~\cite{sak2015fast_interspeech,li2019improving,inaguma2020streaming}.
This is because (1) the model is typically equipped with a time-restricted streaming encoder having limited future contexts, and (2) the model is optimized with an end-to-end objective, which encourages the decoder to use as many future observations as possible.

Previous studies on frame-synchronous models tackled this problem by shifting output frames~\cite{pundak2016lower} and leveraging frame-wise alignment supervision~\cite{sak2015acoustic_asru,dfsmn_ctc,hu2020exploring,mahadeokar2020alignment}.
As for label-synchronous models, Inaguma~{\etal} leveraged alignment information extracted from a hybrid system to reduce the emission latency of MoChA~\cite{inaguma2020streaming}.
However, this approach still depends on a hybrid model and is not a purely end-to-end solution.
While the delayed token generation problem occurs similarly in frame-synchronous models, CTC models are better than AED models in terms of latency, because they are optimized with the forward-backward algorithm and assume conditional independence on a per-frame basis.
Moreover, the peaky alignments learned in CTC are expected to be compatible with the token boundaries in MoChA.

In this article, we propose a novel purely end-to-end training method to enhance the alignment learning process of streaming MoChA models without external alignments.\footnote{Although we focus on MoChA in this work, the proposed method can be applied to any AED model that calculates attention scores.}
We regard the peaks in CTC alignments as a reference for the tokens boundaries in MoChA; thus, we train a MoChA model to mimic a CTC model in order to detect token boundaries at similar positions.
We refer to this method as \textit{CTC synchronous training} (CTC-ST)~\cite{inaguma2020_ctcsync}.
This boundary supervision greatly eases the optimization of MoChA, especially with contaminated inputs as in SpecAugment~\cite{specaugment}.
This is because the accumulation of alignment errors in MoChA can be recovered with the help of the CTC alignments.
The CTC model is jointly optimized with the MoChA model by having them share an encoder to encourage monotonic alignments in the MoChA decoder, similarly to the joint CTC/Attention framework~\cite{kim2017joint}.
In the proposed method, however, the CTC alignments are further provided to the MoChA decoder as supervision of token boundaries to restrict their positions.
Because the alignment knowledge learned in CTC is transferred to improve MoChA's alignment in a unified architecture, we regard this framework as a form of \textit{self-distillation}.\footnote{Unlike conventional knowledge distillation~\cite{hinton2015distilling}, we focus on token boundary positions instead of probability distributions.}

Experimental evaluations on four benchmark datasets show that CTC-ST significantly improves the recognition accuracy, especially for long-form and noisy speech.
We also demonstrate that CTC-ST can reduce the emission latency without external alignment information and achieve a tradeoff of the accuracy and emission latency comparable to that of alignment knowledge distillation from a hybrid system~\cite{inaguma2020streaming}.
Finally, we compare MoChA with RNN-T in recognition accuracy and emission latency to demonstrate that CTC-ST can close the performance gap, so that the best MoChA system achieves recognition accuracy comparable to that of RNN-T and lower emission latency.\footnote{This work is an extension of our previous works~\cite{inaguma2020_ctcsync}. In this article, we add more evaluations regarding various datasets, the emission latency, and the inference speed. Also new are the comparisons with alignment knowledge distillation from a hybrid system~\cite{inaguma2020streaming} and RNN-T.}

\section{Related work}
\subsection{Streaming attention-based encoder-decoder model}
We categorize streaming AED models into two groups in terms of how they segment speech frames for token generation.

\subsubsection{Segmentation on encoder side}
The first method in this category is NT~\cite{neural_transducer,jaitly2016online,sainath2018improving}, which performs label-synchronous decoding on every fixed-size input block and moves to the next block if it detects no additional token boundaries.
ACS~\cite{adaptive_computation_steps} extends the idea to an adaptive segmentation policy based on a halting mechanism~\cite{graves2016adaptive}.
CIF~\cite{cif} further enhances ACS by applying a fine-grained segmentation within the encoder output.
SCAMA~\cite{zhang2020} learns to count the number of tokens to be generated in each fixed input chunk.
Sterpu~{\etal} followed a similar idea~\cite{sterpu2020learning}.
Triggered attention~\cite{moritz2019triggered_icassp2019,moritz2019streaming_asru2019,moritz2020streaming} is based on the joint CTC/Attention framework and truncates encoder outputs with CTC spikes to perform global attention over the past encoder outputs from each spike position, but the decoding complexity is quadratic.
Our work is different in that we leverage CTC alignments only during training, and the decoding complexity is linear.
A scout network~\cite{wang2020low} learns to detect word boundaries with alignment information from a hybrid system during training to reduce the emission latency, and it performs global attention similarly to triggered attention.
However, it introduces a dependency on frame-wise alignment supervision.
These methods can detect token boundaries regardless of the input length, but they are limited by not using contextual information is not used for segmentation.

\vspace{2mm}
\subsubsection{Segmentation on decoder side}
The methods in this category leverage decoder states as a query to segment the input speech for every token.
Local windowing methods were proposed first~\cite{luong2015effective,chan2016online,tjandra2017local,hou2017gaussian}.
GMM attention~\cite{graves2013generating,chiu2019comparison} forces the center of attention to move monotonically to the end of the encoder output.
Kong~{\etal} further incorporated source-side information~\cite{kang2021learning}.
NAT~\cite{luo2017learning} trains stochastic variables with a policy gradient, and the optimization was further improved in~\cite{chiu2017online}.
Hard monotonic attention (HMA)~\cite{hard_monotonic_attention} also introduces stochastic variables to detect token boundaries, but the model can be trained efficiently with a cross-entropy objective.
MoChA~\cite{mocha} relieves the strong monotonic constraint in HMA by introducing additional soft attention over a small window.
Miao~{\etal} proposed stable MoChA (sMoChA) by simplifying the attention calculation in MoChA to ease the optimization~\cite{online_hybrid_ctc_attention}.
They further proposed monotonic truncated attention (MTA)~\cite{online_hybrid_ctc_attention_taslp2020} by using soft attention scores in HMA at test time as well to reduce the gap between training and testing.
HMA was further extended to the Transformer decoder~\cite{tsunoo2019towards,inaguma2020enhancing}, while Li~{\etal} extended the idea of ACS to the Transformer decoder~\cite{li2020transformer}.
Incremental decoding uses offline models for streaming applications, but the decoding complexity is quadratic~\cite{novitasari2019sequence,liu2020,nguyen2020super,tsunoo2020streaming}.

\subsection{Emission latency in E2E ASR model}
Emission latency inevitably occurs in any E2E ASR model because sequence-level optimization allows the model to use as much future information as possible.
This problem was tackled in CTC acoustic models for the first time by constraining CTC paths during marginalization with frame-level alignments from a hybrid system~\cite{sak2015learning_icassp}.
Zhang~{\etal} trained a CTC model jointly with a frame-level cross-entropy~\cite{dfsmn_ctc}.
Similar methods have also been investigated for RNN-T by pretraining the encoder with a frame-level cross-entropy~\cite{hu2020exploring} and applying joint training~\cite{mahadeokar2020alignment}.
Inaguma~{\etal} investigated the application of alignment information to MoChA and reduced the recognition errors and emission latency simultaneously~\cite{inaguma2020streaming}.
However, such alignment information is not necessarily available.
Recently, FastEmit~\cite{yu2021fastemit} was proposed by designing a new training objective for RNN-T to reduce the emission latency without any frame-level supervision, which was applied to two-pass E2E architectures in the voice search task successfully~\cite{li2021better,narayanan2021cascaded}.
The idea of FastEmit was extended to MoChA in~\cite{inaguma2021stableemit}, referred to as StableEmit.
Self-alignment from RNN-T was also leveraged in~\cite{kim2021reducing}.
Yu~{\etal} trained a single RNN-T in both offline and streaming modes (\textit{dual-mode ASR}) by sharing parameters~\cite{yu2021dualmode}.

\subsection{Knowledge distillation for streaming ASR}
Knowledge distillation~\cite{hinton2015distilling} has also been investigated to improve the performance of streaming E2E ASR models.
Previous works focused on distilling knowledge from an offline or streaming teacher model to a weaker streaming student model within the same decoder topology~\cite{takashima2018investigation,kim2018improved,huang2018knowledge,pang2018compression,kurata2018improved,mun2019sequence,kurata2020knowledge,panchapagesan2020efficient}.
In frame-synchronous models, however, there exists a problem that the timing to emit tokens can differ between the teacher and student models, depending on the future context size in the encoder.
Through an approach using bidirectional long short-term memory (BLSTM), Kurata~{\etal} trained BLSTM-CTC to mimic LSTM-CTC in order to generate posterior spikes at similar positions and then distill the frame-level posterior probabilities to the LSTM-CTC model~\cite{kurata2019guiding}.
The idea was extended to RNN-T in~\cite{kurata2020knowledge}.
Ding~{\etal} jointly trained multiple teacher CTC models to synchronize their posterior spikes~\cite{ding2020improving}.
Distillation between different decoder topologies has also been investigated.
Moriya~{\etal} distilled knowledge from a teacher AED model to a student CTC model~\cite{moriya2020distilling}.
Self-distillation in a single E2E ASR model has also been proposed as an in-place operation, from an offline mode to a streaming mode~\cite{tsunoo2020streaming,yu2021dualmode}, and from a Transformer decoder to a CTC layer~\cite{moriya2020self}.
Unlike those previous methods, we focus on distilling the positions of token boundaries learned in a CTC model to an AED model, rather than distilling the posterior distributions.
Moreover, the teacher and student models share the same encoder and are trained jointly from scratch.

\section{Basics}
In this section, we review HMA and MoChA.
Let $\speech=(x_{1},\ldots,x_{\xmax})$ be an input speech sequence, ${\bm h}=(h_{1},\ldots,h_{\hmax})$ be encoder outputs ($\hmax \leq \xmax$), and $\transcript=(y_{1},\ldots,y_{\ymax})$ be the corresponding output token sequence.
The encoder performs downsampling to reduce the input sequence length from $\xmax$ to $\hmax$.
We use $i$ and $j$ as the time indices of the output and input sequences, respectively.

\subsection{Hard monotonic attention (HMA)}\label{ssec:hard_monotonic_attention}
Standard offline AED models are based on the global attention mechanism~\cite{chorowski2015attention}, in which relevant source information is selected according to the target context via attention scores calculated by normalizing energy activations over $\encout$.
However, this prevents the model from performing online streaming recognition, because the decoder must see all the encoder outputs to generate the initial token.
Moreover, the decoding complexity at each generation step is in proportion to the encoder output length $\hmax$.
This results in a total decoding complexity $\mathcal{O}(\hmax \ymax)$.
To start generating tokens when given partial acoustic observations during inference, HMA introduces discrete binary decision processes.
As a result, it can perform decoding with linear-time complexity $\mathcal{O}(\hmax)$ during inference, but it behaves differently between the training and test times.

At test time, the decoder scans the encoder outputs $h_{1}, \ldots, h_{\hmax}$ from left to right.
At every input frame index $j$, the decoder has the option to (1) stop at the current frame $j$ to generate a token or (2) move forward to the next frame $j + 1$ according to a selection probability $\pij \in [0,1]$.
A discrete decision $\zij\in\{0,1\}$ on whether to stop at the $j$-th frame is sampled from a Bernoulli random variable parameterized by $\pij$ as
\begin{eqnarray}
\eij &=& \mbox{MonotonicEnergy}(h_{j}, s_{i}), \label{eq:e_mono} \\
\pij &=& \sigma(\eij), \label{eq:sigmoid} \\
\zij &\sim& \mbox{Bernoulli}(\pij), \nonumber
\end{eqnarray}
where $\eij$ is a monotonic energy activation, $s_{i}$\footnote{In this formulation, $s_{i}$ is updated before calculating $\eij$, unlike in~\cite{hard_monotonic_attention}.} is the $i$-th decoder state, and $\sigma$ is a logistic sigmoid function.
When $\zij=1$, {\ie}, $\pij \geq 0.5$, the decoder stops at an index $j=t_{i}$ (referred to as the \textit{token boundary} of the $i$-th token).
Then, only the corresponding single encoder output $h_{t_{i}}$ is used for generating the next token and updating the decoder state.
The next token boundary $t_{i+1}$ is determined by resuming scanning from the previous token boundary $j=t_{i}$.

However, this hard assignment of $\zij$ is not differentiable.
To perform the standard backpropagation training, the expected alignment scores $\alphaij$ are calculated by marginalizing over all possible alignment paths as follows:
\begin{eqnarray}
\alphaij &=& \pij \sum_{k=1}^{j}\bigg(\alpha_{i-1,k}\prod_{l=k}^{j-1}(1-p_{i,l})\bigg) \nonumber \\
&=& \pij \bigg((1-p_{i,j-1})\frac{\alpha_{i,j-1}}{p_{i,j-1}}+\alpha_{i-1,j}\bigg). \label{eq:monotonic_attention_alpha}
\end{eqnarray}
Because $\alphaij$ in Eq.~\eqref{eq:monotonic_attention_alpha} introduces a recurrence relation, it is difficult to calculate it in parallel over the input indices.
However, by substituting $q_{i,j}=\alphaij / \pij$, it can be calculated efficiently with the cumulative sum and product operations, denoted respectively as {\cumsum} and {\cumprod}, as follows:
\small
\begin{eqnarray*}
{\bm \alpha}_{i,:}={\bm p}_{i} \cdot \mbox{\cumprod}(1-{\bm p}_{i,:}) \cdot \mbox{\cumsum}\bigg(\frac{{\bm \alpha}_{i-1,:}}{\mbox{\cumprod}(1-{\bm p}_{i,:})}\bigg).
\end{eqnarray*}
\normalsize
Finally, the monotonic energy activation $\eij$ in Eq.~\eqref{eq:e_mono} is implemented as
\small
\begin{eqnarray*}
{\rm MonotonicEnergy}(h_{j}, s_{i}) &=& g\frac{v^{\mathsf T}}{||v||}f({\mathbf W}_{\rm h}h_{j} + {\mathbf W}_{\rm s}s_{i} + b) + r,
\end{eqnarray*}
\normalsize
where $f$ is the nonlinear activation, $g$ and $v$ are parameters for weight normalization, and ${\mathbf W}_{\rm h}$, ${\mathbf W}_{\rm s}$, $b$, and $r$ are trainable parameters.
We use a rectified linear unit (ReLU) activation function as $f$.
Following~\cite{hard_monotonic_attention}, the scalar offset parameter $r$ is initialized as -4 in this work.
To ensure the discreteness of $\pij$, a zero-mean, unit-variance Gaussian noise is added to the pre-sigmoid activation in Eq.~\eqref{eq:sigmoid} during training.
The subsequent token generation processes are the same as in the global AED model.

\subsection{Monotonic chunkwise attention (MoChA)}\label{ssec:mocha}
In HMA, source information is restricted to a single encoder output, and this strong constraint greatly sacrifices the accuracy in general.
To overcome this problem by leveraging the surrounding contexts, MoChA introduces an additional soft attention mechanism over a fixed window of width $\chunkmocha$ on top of HMA.

At test time, soft attention scores $\betaij$ are calculated over $\chunkmocha$ encoder outputs from every token boundary $t_{i}$ as follows:
\begin{gather}
\betaij=\exponential(\uij)/\sum_{l=t_{i} - \chunkmocha + 1}^{t_{i}}{\exponential(u_{i,l})}, \label{eq:mocha_beta_test} \\
c_{i}=\sum_{j=t_{i} - \chunkmocha + 1}^{t_{i}} \betaij h_{j}, \nonumber
\end{gather}
where $\uij$ is the chunk energy activation formulated as in Eq.~\eqref{eq:e_mono} without weight normalization and the offset parameter $r$, and $c_{i}$ is a context vector for the $i$-th token.

During training, $\betaij$ can be calculated on top of the expected alignment score $\alphaij$ as
\begin{gather}
\betaij=\sum_{k=j}^{j + \chunkmocha - 1}\bigg(\alpha_{i,k}\exponential(\uij)/\sum_{l=k - \chunkmocha + 1}^{k}{\exponential(u_{i,l})}\bigg). \label{eq:mocha_beta_train}
\end{gather}
The computation of $\betaij$ is expensive because of the nested summation.
Fortunately, it can be computed more efficiently with a moving sum operation, denoted as {\movsum}, as follows:
\begin{gather*}
\betaij=\exponential(u_{i,:}) \cdot \mbox{\movsum}(\frac{\alpha_{i,:}}{\mbox{\movsum}(\exponential(u_{i,:},\chunkmocha,1))},1,\chunkmocha), \\
\mbox{\movsum}(\speech,b,f)_{n}=\sum_{m=n-(b-1)}^{n+f-1}x_{m}.
\end{gather*}
This computation can be implemented by 1-dimensional convolution.
$c_i$ is calculated as $c_{i} = \sum_{j=1}^{\hmax} \betaij h_{j}$.
The objective function is formulated as the negative log-likelihood $\lossmocha=- \log \probmocha(\transcript|\speech)$, where $\probmocha$ is the output probability distribution of MoChA.

To encourage MoChA to learn monotonic alignments, we train it jointly with an auxiliary CTC objective $\lossctc=- \log{\probctc(\transcript|\speech)}$, where $\probctc$ is the CTC output probability distribution, by sharing the encoder sub-network~\cite{kim2017joint,hybrid_ctc_attention}.
Moreover, to avoid vanishing of $\alphaij$, quantity loss $\lossqua$ is introduced by making the expected total number of token boundaries closer to the reference output sequence length $\ymax$ as follows~\cite{cif,inaguma2020streaming}:
\begin{eqnarray}
\lossqua=|\ymax - \sum_{i=1}^{\ymax}{\sum_{j=1}^{\hmax}{\alphaij}}|. \label{eq:quanity_loss}
\end{eqnarray}
We refer to this technique as \textit{quantity regularization} (QR).\footnote{QR was investigated for DeCoT in~\cite{inaguma2020streaming} for the voice search task, and it did not bring any improvement for a naive MoChA model.
However, we found that it gave large improvements for the naive model in our experiments.
We attribute this to the longer input sequence length distributions and the smaller training data sizes of the corpora used in this work.
Therefore, unless otherwise noted, we use QR for the baseline MoChA models in this work.}
Note that any external alignment information is not used here.
The total objective function $\losstotal$ is defined as a linear interpolation of $\lossmocha$, $\lossctc$, and $\lossqua$ as follows:
\begin{eqnarray}
\losstotal=(1- \lambdactc) \lossmocha + \lambdactc \lossctc + \lambdaqua \lossqua, \label{eq:total_loss_mocha_baseline}
\end{eqnarray}
where $\lambdactc$ ($0 \le \lambdactc \le 1$) and $\lambdaqua$ ($\ge 0$) are tunable weights for the CTC loss and quantity loss, respectively.

\section{Problems in MoChA}\label{sec:optimization_problem}
In this section, we review two major problems in MoChA: vanishing alignment probabilities and delayed token generation.

\subsection{Vanishing alignment probabilities}\label{ssec:optimization_problem}
Conventional alignment models such as the hidden Markov model (HMM)~\cite{gales2008application} and CTC models use the forward-backward algorithm to calculate alignment probabilities.
However, it is not straightforward to apply that algorithm to the HMA mechanism.
This is because HMA does not normalize the monotonic energy $\eij$ across the entire set of encoder outputs $\encout$ to obtain the expected alignment score $\alphaij$, which means that $\alphaij$ is not a valid probability.
Moreover, the decoder is an autoregressive model, and an incremental left-to-right update of the decoder state is required for each token.
Therefore, during the marginalization process at training time, $\alphaij$ depends only on past alignments, as can be seen in Eq.~\eqref{eq:monotonic_attention_alpha}, and it can quickly be attenuated as the number of decoding steps increases~\cite{online_hybrid_ctc_attention_taslp2020,he2019robust}, because $\sum_{j=1}^{\hmax}{\alphaij} < 1$.
This is especially problematic for long-form speech because the model is more likely to fail to learn the scale of $\pij$ properly in the latter steps.
Accordingly, the gap in the HMA behaviors between training and testing is widened.
This leads to premature endpointing, which increases deletion errors~\cite{hsiao2020online,zhao2021preventing}.

\begin{figure}[t]
  \centering
  \includegraphics[width=0.99\linewidth]{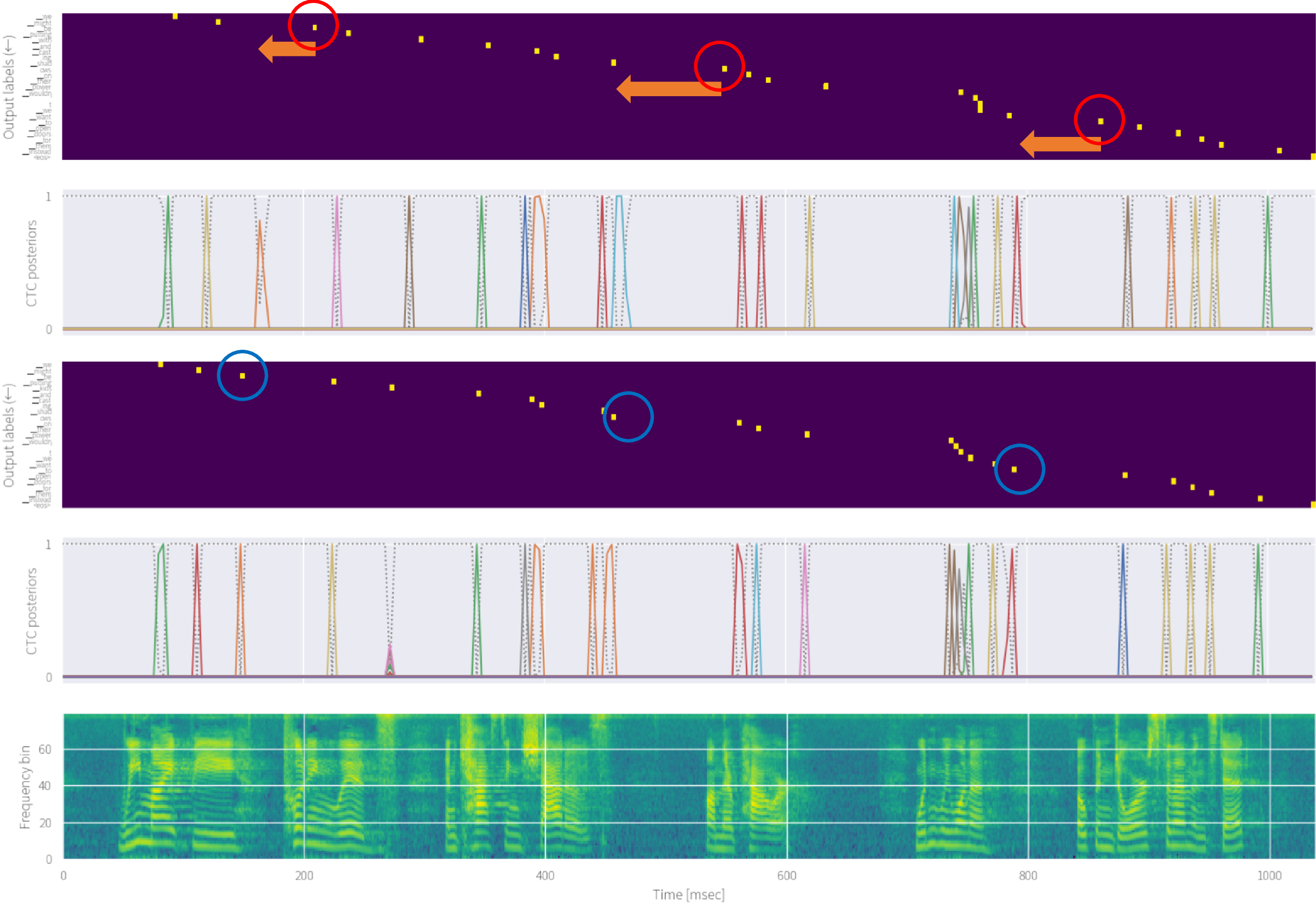}
  \caption{Visualization of the token boundaries (yellow dots) of the baseline UniLSTM - MoChA model (top, \textit{T5} in Table~\ref{tab:taslp2021_result_tedlium2}) and the proposed model with CTC-ST (bottom, \textit{T6}). Reference: ``\textit{we might be putting lids and casting shadows on their power wouldn't we want to open doors for them instead}.''}
  \label{fig:taslp2021_attention_plot}
\end{figure}

\subsection{Delayed token generation}\label{ssec:delayed_token_generation_problem}
To enable online inference, a streaming ASR model needs to be equipped with a time-restricted encoder, which does not use enough future information.
However, because E2E models are optimized with sequence-level criteria, {\ie}, the cross-entropy, future encoder outputs are used as much as possible~\cite{sak2015acoustic_asru,adaptive_mocha,kim2020attention,inaguma2020streaming}.
As a result, token boundaries are shifted several frames ahead from the actual acoustic boundary positions.
Moreover, MoChA allows emission of multiple tokens at the same input index.
These problems inevitably cause a large perceived latency and make the model unusable in online ASR.

Figure~\ref{fig:taslp2021_attention_plot} (top) shows an example of token boundaries emitted from a baseline MoChA model trained with an auxiliary CTC loss.
The yellow dots and the spikes below represent the token boundaries from MoChA and the CTC posterior probabilities, respectively.
It is known that the posterior probabilities in a well-trained CTC model tend to peak in sharp spikes~\cite{ctc_graves}.
In the figure, we can observe that MoChA's token boundaries are indeed shifted to the right (future side) from the actual acoustic boundaries and are poorly aligned to the CTC spikes.

\section{Alignment knowledge distillation from hybrid ASR system}\label{sec:akd_hybrid_asrt}
This section describes previous approaches to tackle the delayed token generation problem in MoChA by leveraging word alignments extracted from a hybrid ASR system~\cite{inaguma2020streaming}.
We review two methods in~\cite{inaguma2020streaming}: \textbf{De}lay \textbf{Co}nstrained \textbf{\xmax}raining (DeCoT) and \textbf{Min}imum \textbf{L}atency \textbf{\xmax}raining (MinLT), which can reduce both the emission latency and recognition errors.
Because the alignment knowledge bootstrapped in a hybrid system is transferred to an AED model, we refer to the procedure as \textit{alignment knowledge distillation} from the hybrid system.

Let $\alignment=(a_{1}, \cdots, a_{\xmax})$ ($a_{j}$: $\vocab$-dimensional one-hot vector; $\vocab$: vocabulary size of the hybrid system) be a frame-level word alignment corresponding to the input sequence $\speech$, and let $\bref=({\rm b}^{\rm ref}_{1}, \cdots, {\rm b}^{\rm ref}_{\ymax})$ be a sequence of endpoints of token boundaries for a reference transcription $\transcript=(y_{1}, \cdots, y_{\ymax})$.
To convert the word alignment to a subword alignment compatible with MoChA, we divide the total time duration of each word in $\alignment$ by the ratio of the character length of each subword.
Finally, we select the end timestamp as the token boundary.

\subsection{Delay constrained training (DeCoT)}\label{ssec:decot}
In DeCoT, inappropriate alignment paths that poorly match the reference alignment are removed by masking out their scores $\alphaij$ during marginalization.
Then, $\alphaij$ in Eq.~\eqref{eq:monotonic_attention_alpha} is reformulated to $\alphadecotij$ as follows:
\small
\begin{eqnarray*}
\alphadecotij=\begin{cases}
     \pij \bigg((1-p_{i,j-1})\frac{\alpha_{i,j-1}}{p_{i,j-1}}+\alpha_{i-1,j}\bigg) & (j \le \drefi + \deltadecot) \\
     0 & ({\rm otherwise}),
\end{cases}
\end{eqnarray*}
\normalsize
where $\deltadecot$ is a hyperparameter to control the acceptable delay, and $\drefi$ is a reference boundary of the $i$-th token transferred from the hybrid system.

As explained in Section~\ref{ssec:optimization_problem}, MoChA has a problem of an exponential decay of $\alphaij$, and the masking for $\alphadecotij$ accelerates it further.
To recover the proper scale of $\alphadecotij$, $\lossqua$ in Eq.~\eqref{eq:quanity_loss} is calculated with $\alphadecotij$ instead of $\alphaij$ as
\begin{eqnarray}
\lossquadecot=|\ymax - \sum_{i=1}^{\ymax}{\sum_{j=1}^{\hmax}{\alphadecotij}}|. \label{eq:quantity_loss_decot}
\end{eqnarray}
Intuitively, QR in DeCoT emphasizes the valid alignment paths during marginalization.
Moreover, this also leads to better estimation of $\betaij$ in Eq.~\eqref{eq:mocha_beta_train}.
Accordingly, the total objective function in Eq.~\eqref{eq:total_loss_mocha_baseline} is modified as follows:
\begin{eqnarray}
\losstotal=(1- \lambdactc) \lossmocha + \lambdactc \lossctc + \lambdaqua \lossquadecot. \label{eq:total_loss_decot}
\end{eqnarray}
Unlike in~\cite{inaguma2020streaming}, we also add a CTC loss as an auxiliary objective for DeCoT.

\subsection{Minimum latency training (MinLT)}\label{ssec:minlt}
While DeCoT can effectively reduce the emission latency, a fixed buffer size $\deltadecot$ must be predefined for every token.
However, the emission latency can vary depending on the speaking rate, subwords, and so on.
To reduce the latency of each token more flexibly, MinLT directly minimizes the expected latency so that the expected token boundaries in MoChA get closer to the corresponding reference boundaries.
A differentiable expected latency loss $\lossminlt$ is specified as
\begin{eqnarray}
\bmochai &=& \sum_{j=1}^{\hmax}j \cdot \alphaij, \\
\lossminlt &=& \frac{1}{\ymax}\sum_{i=1}^{\ymax}{|\drefi - {\rm b}_{i}^{\rm mocha}|}, \label{eq:minlt_loss}
\end{eqnarray}
where $\bmochai$ is the expected boundary position in MoChA for the $i$-th token during training.
The total objective function in Eq.~\eqref{eq:total_loss_mocha_baseline} is modified as follows:
\begin{eqnarray}
\losstotal=(1- \lambdactc) \lossmocha + \lambdactc \lossctc + \lambdaminlt \lossminlt, \label{eq:total_loss_minlt}
\end{eqnarray}
where $\lambdaminlt$ ($\ge 0$) is a hyperparameter to control the latency.

\section{Proposed method: CTC synchronous training}\label{sec:ctc_st}
In this section, we propose a novel training method to alleviate the problems of vanishing alignment probabilities and delayed token generation in MoChA by leveraging CTC alignments as a reference for token boundaries.

\begin{figure}[t]
  \centering
  \includegraphics[width=0.99\linewidth]{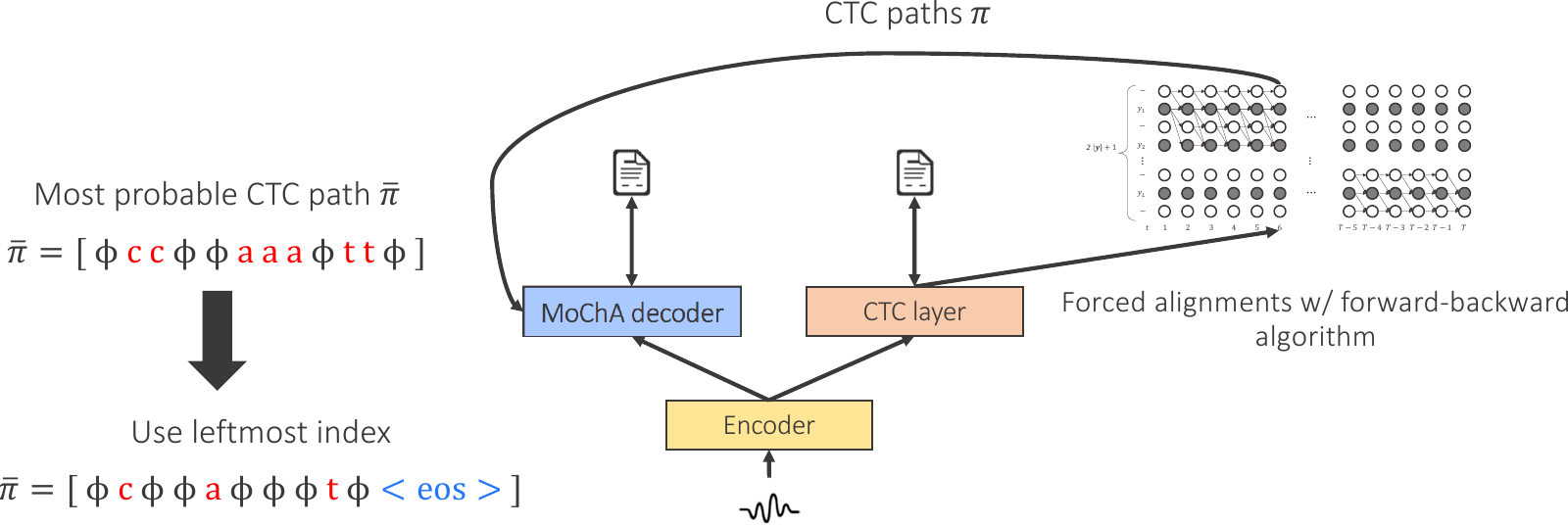}
  \caption{Overview of CTC synchronous training (CTC-ST)}
  \label{fig:taslp2021_overview}
\end{figure}

\subsection{Overview}
As observed in Fig.~\ref{fig:taslp2021_attention_plot}, there exists a gap in the timing to emit tokens between the CTC and MoChA models even when they share the same encoder.
However, we can see that the CTC spikes are closer to the actual acoustic boundaries.
The reason is that CTC does not suffer from alignment error propagation because of the optimization with the forward-backward algorithm and the assumption of conditional independence on a per-frame basis.
Therefore, we expect that CTC can generate more reliable alignments than MoChA and serve as an effective guide for MoChA to learn to detect token boundaries more accurately.

Motivated by this reasoning, we propose \textit{CTC synchronous training} (CTC-ST), in which a MoChA model is trained to mimic a CTC model in order to generate token boundaries at similar positions.
Figure~\ref{fig:taslp2021_overview} shows a system overview.
Both the MoChA and CTC branches are jointly optimized by sharing the encoder, and reference token boundaries are obtained from the CTC alignments generated from the CTC branch.
Therefore, the CTC model serves not only as a regularizer to enhance the encoder representations but also as a teacher alignment model to estimate accurate token boundary positions.
In this sense, we regard CTC-ST as a form of \textit{self-distillation} from CTC to MoChA.
Specifically, the synchronization of token boundaries in CTC-ST can be viewed as explicit interaction between MoChA and CTC models on the \textit{decoder} side, unlike in the conventional joint CTC/Attention framework~\cite{kim2017joint}.
However, we only leverage the discrete token boundary positions rather than the alignment probability distributions as in the conventional knowledge distillation method~\cite{hinton2015distilling}.

Because CTC is not allowed to emit multiple symbols at the same input index, the reference token boundaries from the CTC alignments can also enforce the monotonicity of $\alphaij$ in MoChA, which leads to emission latency reduction as well.
Moreover, unlike the methods described in Section~\ref{sec:akd_hybrid_asrt}, the entire model can be trained in an end-to-end manner without relying on external alignments extracted from a hybrid system or manual annotation.

\subsection{Extraction of CTC alignments}\label{ssec:alignment_extraction}
We use the most probable CTC path $\bar{\ctcpath}=\argmax_{\ctcpath}p(\ctcpath | \speech)$ ($|\bar{\ctcpath}|=\hmax$), given by Viterbi alignment, via forced alignment with the forward-backward algorithm in a manner similar to triggered attention~\cite{moritz2019triggered_icassp2019}.
The time indices of non-blank tokens in $\bar{\ctcpath}$ are used as the reference token boundaries $\bctc=({\rm b}^{\rm ctc}_{1},\ldots,{\rm b}^{\rm ctc}_{\ymax})$.
When repeated non-blank labels exist, the leftmost index corresponding to the same non-blank token is used as a reference token boundary.
The last time index $\hmax$ is used for the end-of-sentence (EOS) mark, $\langle$eos$\rangle$.
For instance, given a CTC path $\bar{\ctcpath}=$\texttt{[$\blank$,c,c,$\blank$,a,a,a,$\blank$,t,t,$\blank$]} ($\blank$: blank) corresponding to a reference transcription ``\texttt{c a t $\langle$eos$\rangle$}'', we convert it to \texttt{[$\blank$,c,$\blank$,$\blank$,a,$\blank$,$\blank$,$\blank$,t,$\blank$,$\langle$eos$\rangle$]} and then extract the time indices of the non-blank tokens $\bctc=$\texttt{(2,5,9,11)} (1-indexed).
Unless otherwise specified, $\bctc$ is generated with the model parameters at each training step \textit{on the fly} and is expected to get more accurate as the training continues.
We can also pre-compute $\bctc$ and use the fixed boundaries throughout training when adopting curriculum learning; this approach is described in Section~\ref{ssec:curriculum_learning} and analyzed in Section~\ref{ssec:precompute_alignment}.

\subsection{Optimization}
We define the objective function of CTC-ST $\losssync$ (hereafter, the CTC-ST loss) as
\begin{eqnarray}
\losssync=\frac{1}{\ymax} \sum_{i=1}^{\ymax}|\bctci - \bmochai|. \label{eq:ctc_st_loss}
\end{eqnarray}
The total objective function in Eq.~\eqref{eq:total_loss_mocha_baseline} is reformulated as
\begin{eqnarray}
\losstotal=(1- \lambdactc) \lossmocha + \lambdactc \lossctc + \lambdasync \losssync, \label{eq:total_loss_ctcst}
\end{eqnarray}
where $\lambdasync$ ($\ge 0$) is a tunable hyperparameter.
When using CTC-ST, we do not use the quantity loss and CTC-ST loss simultaneously.
This is because we found that the combination was not effective in our experiments, as described in Section~\ref{sec:taslp2021_results}.
Instead, we propose an effective curriculum learning strategy in Section~\ref{ssec:curriculum_learning}.

\subsection{Curriculum learning strategy}\label{ssec:curriculum_learning}
To calculate the effective gradient via Eq.~\eqref{eq:ctc_st_loss} in CTC-ST, it is necessary to estimate reasonable expected token boundary positions in MoChA during training.
However, $\alphaij$ tends to diffuse over several frames in the early training stage.
Again, we note that $\alphaij$ is not explicitly normalized to sum up to one.
Moreover, CTC alignments would not be very accurate in the early training stage either.
Therefore, applying CTC-ST from a random parameter initialization leads to unstable, slower convergence.
To tackle this problem, we propose a simple but effective curriculum learning strategy composed of the following two stages.
\vspace{2mm}
\subsubsection{Stage 1}
We first train a MoChA model equipped with a bidirectional encoder ({\eg}, BLSTM) together with QR by applying Eq.~\eqref{eq:total_loss_mocha_baseline} from scratch until convergence.
As the bidirectional encoder can see the entire context, we refer to this model as ``offline.''
In this stage, we expect the model to learn a proper scale of $\alphaij$.

\vspace{2mm}
\subsubsection{Stage 2}
Next, we optimize a MoChA model equipped with a latency-controlled bidirectional encoder (LC-BLSTM)~\cite{latency_controlled_blstm} with CTC-ST by applying CTC-ST with Eq.~\eqref{eq:total_loss_ctcst}.
We initialize the parameters with values optimized in stage 1.
Because the BLSTM and LC-BLSTM encoders have the same model structure, we can reuse all of the parameters; the only difference between them is the lookahead context size.
The optimizer's parameters and the learning rate are reset at the beginning of stage 2.
In this stage, we expect the model to learn accurate token boundary location.
When using a unidirectional LSTM encoder, the same encoder is used in both stages.
We also apply this curriculum learning strategy to DeCoT and MinLT to stabilize the training~\cite{inaguma2020streaming}.

\subsection{Combination with SpecAugment}\label{ssec:combine_specaug}
Recently, SpecAugment~\cite{specaugment} has shown the capability to greatly improve the performance of E2E ASR models.
SpecAugment is an on-the-fly data augmentation method that introduces stochastic time and frequency masks into input speech.
$M_{\xmax}$ time masks, whose size is sampled from a uniform distribution ${\cal U}(0,T)$ are applied to the input log-mel spectrogram.
Similarly, frequency masks are also applied with mask parameters $M_{F}$ and $F$.
However, such input masks easily collapse the recurrence of $\alphaij$ in Eq.~\eqref{eq:monotonic_attention_alpha} right after the masked region.
Although such a problem does not exist in offline global AED models or frame-synchronous models, it is a severe problem in MoChA.
In our experiments, in fact, the performance of the baseline MoChA model was degraded by applying SpecAugment.

In contrast, CTC-ST can help MoChA recover the collapsed $\alphaij$ in the masked region by leveraging the CTC spikes because CTC assumes conditional independence on a per-frame basis.
Therefore, we expect that CTC-ST is beneficial for MoChA to learn monotonic alignments that are robust for noisy inputs.

\section{Experimental evaluation}\label{sec:exp}
\subsection{Datasets}\label{ssec:corpus}
\begin{figure}[t]
  \centering
  \includegraphics[width=0.90\linewidth]{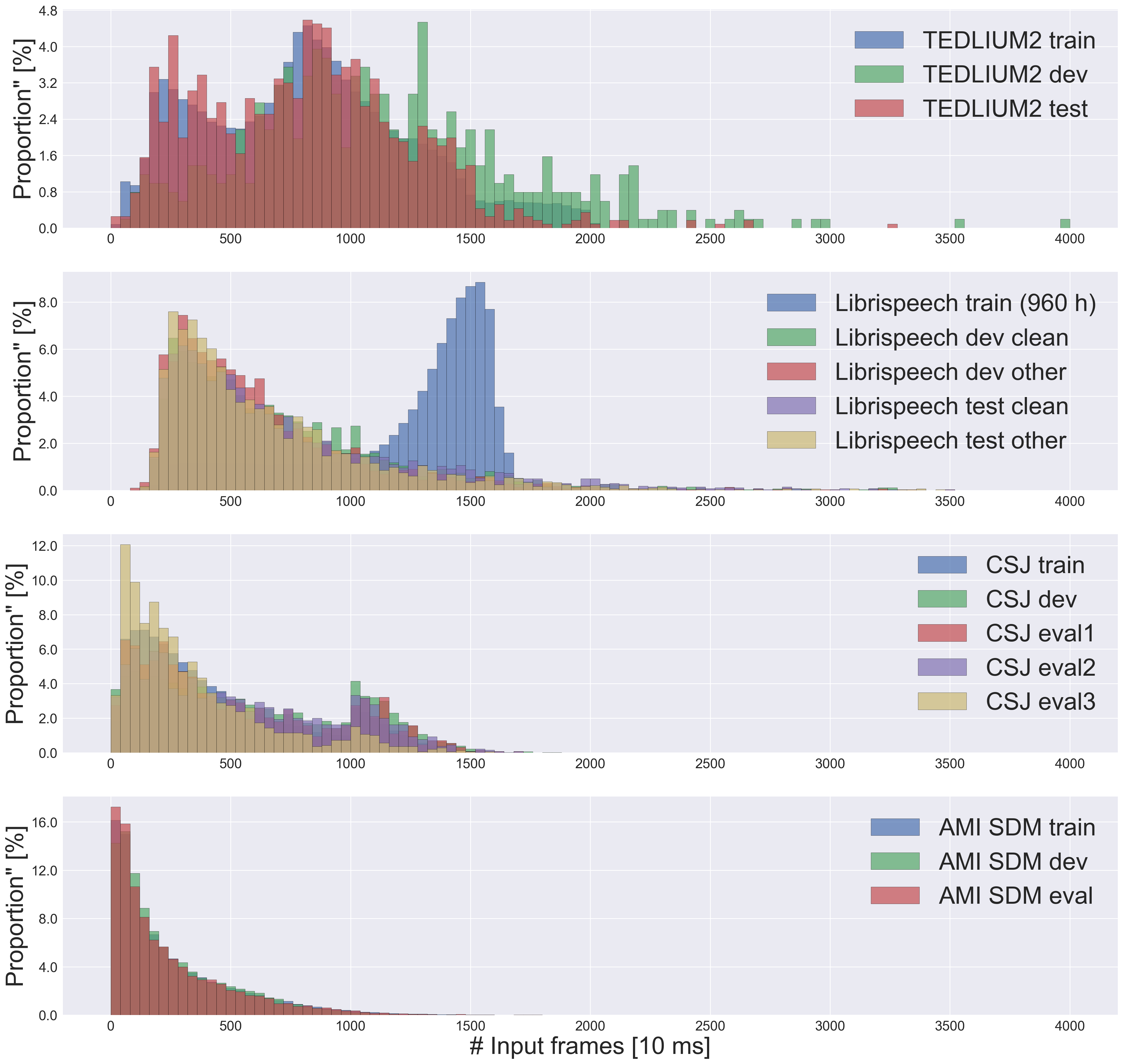}
  \caption{Utterance length distributions in the TEDLIUM2, Librispeech, CSJ, and AMI (SDM) corpora}
  \label{fig:taslp2021_corpus_utt_dist}
\end{figure}

We used the TEDLIUM release 2 (TEDLIUM2)~\cite{tedlium} and Librispeech~\cite{librispeech}, the Corpus of Spontaneous Japanese (CSJ)~\cite{csj}, and the single distant microphone (SDM) portion of the AMI Meeting Corpus~\cite{ami} for our experimental evaluations.
The corpus statistics and the utterance length distributions are presented in Table~\ref{tab:corpus} and Fig.~\ref{fig:taslp2021_corpus_utt_dist}, respectively.
We used 10k vocabularies based on the byte pair encoding (BPE) algorithm~\cite{sennrich2015neural} except for the AMI corpus, for which a vocabulary of 500 BPE units was used.

\begin{table}[t]
    \centering
    \begingroup
    \caption{Corpus statistics}\label{tab:corpus}
    \scalebox{0.98}{
    \begin{tabular}{@{}ccccc@{}} \toprule
    Corpus & Hours & Domain & Language & BPE size \\ \midrule
    TEDLIUM2~\cite{tedlium} & 210 & Lecture & English & 10k \\
    Librispeech~\cite{librispeech} & 960 & Reading & English & 10k\\
    CSJ~\cite{csj} & 586 & Lecture & Japanese & 10k \\
    AMI (SDM)~\cite{ami} & 100 & Meeting & English & 500 \\
    \bottomrule
    \end{tabular}
    }
    \endgroup
\end{table}

\subsection{Experimental configuration}
Using the Kaldi toolkit~\cite{kaldi}, we extracted 80-channel log-mel filterbank coefficients computed with a 25-ms window that was shifted every 10ms.
\textcolor{\colorblue}{
Input features were normalized by the global mean and variance calculated on each training set.}
We removed utterances longer than 16 seconds from the training data to conserve the GPU memory capacity.
The training utterances were sorted by their input lengths in ascending order during the entire training stage.
We applied 3-fold speed perturbation~\cite{speed_perturbation} to the TEDLIUM2 and AMI corpora with factors of 0.9, 1.0, and 1.1.

The encoders consisted of two CNN blocks followed by five layers of (LC-)BLSTM or unidirectional LSTM.
Each CNN block consisted of two CNN layers, each of which had a $3 \times 3$ filter followed by a max-pooling layer with a stride of $2 \times 2$.
This resulted in a 4-fold frame rate reduction in total and introduced a 60ms lookahead latency for every output of the CNN blocks.
We set the number of units in each (LC-)BLSTM layer to 512 per direction.
To reduce the input dimension of the subsequent (LC-)BLSTM layer, we summed the LSTM outputs in both directions at every layer~\cite{tuske2019advancing}.
For a unidirectional LSTM encoder, the unit size was increased to 1024.
In this article, we denote an LC-BLSTM encoder with a hop size of $\blocksizecurrent$ frames and a future context of $\blocksizeright$ frames as ``LC-BLSTM-$\blocksizecurrent$+$\blocksizeright$.''
The decoder was a single layer of unidirectional LSTM with 1024-dimensional units.
We set the window size $\chunkmocha$ of chunkwise attention in MoChA to 4.
Offline global AED models used the location-based attention~\cite{chorowski2015attention}.

We also trained RNN-T models with the same encoder for comparison.
The RNN-T models had a two-layer LSTM prediction network with 1024 memory units and a joint network with 512 units.
\textcolor{\colorblue}{
The 1k BPE was used for the vocabulary except for 500 units on the AMI corpus.
These vocabulary sizes were selected to achieve the best performance for RNN-T.}
We also used an auxiliary CTC loss for RNN-T as $\losstotal=(1 - \lambdactc) {\cal L}_{\rm rnnt} + \lambdactc \lossctc$, where ${\cal L}_{\rm rnnt}$ is a RNN-T loss.
$\lambdactc$ was set to 0.3.

The Adam optimizer~\cite{adam} was used with an initial learning rate of $1e-3$, which was then decayed exponentially.
For the TEDLIUM2 and AMI corpora, 4k warmup steps were used.
We applied dropout and label smoothing~\cite{label_smoothing} with probabilities of 0.4 and 0.1, respectively.
The weight of the quantity loss $\lambdaqua$ in Eq.~\eqref{eq:total_loss_mocha_baseline} was set to 2.0, \textcolor{\colorblue}{0.1}, 1.0, and 1.0 for the TEDLIUM2, Librispeech, CSJ, and AMI corpora, respectively.
$\lambdaqua$ in Eq.~\eqref{eq:total_loss_decot} was set to 2.0.
The CTC-ST weight $\lambdasync$ in Eq.~\eqref{eq:total_loss_ctcst} was set to 1.0 for all corpora, unless otherwise noted.
$\lambdactc$ was set to 0.3 in all models.
All training was performed with a single GPU.

For inference, we used a 4-layer LSTM LM with 1024 units per layer.
For AED models, we used a beam width of 10 and scores normalized by the output sequence length at every output timestep, except for global AED models on the AMI corpus\footnote{This was because the utterance lengths in the AMI corpus are relatively short.}.
Joint CTC decoding was performed for global AED models~\cite{hybrid_ctc_attention}.
For RNN-T models, we used the breadth-first time-synchronous decoding (TSD) algorithm with a monotonic constraint (hereafter, \textit{mono-TSD})~\cite{tripathi2019monotonic} to speed up decoding\footnote{Although we did not explicitly use a monotonic RNN-T loss, the joint CTC training enforced a similar effect.}, and we merged paths corresponding to the same label history, except for the AMI corpus\footnote{Again, this was because the utterance lengths are relatively short.}.
We also reduced the beam width of streaming RNN-T models to 5 because of the inference speed constraint.
Our implementation is publicly available.\footnote{\url{https://github.com/hirofumi0810/neural_sp}}

\begin{table}[t]
    \centering
    \begingroup
    \caption{Results without SpecAugment on \underline{TEDLIUM2}. QR: quantity regularization. CTC-ST: CTC-synchronous training.}\label{tab:taslp2021_result_tedlium2}
    \scalebox{1.0}{
    \begin{tabular}{@{}lccc@{}} \toprule
     \multirow{2}{*}{Model} & \multirow{2}{*}{Seed} & \multicolumn{2}{c}{WER [$\%$] ($\downarrow$)} \\ \cmidrule(lr){3-4}
      & & dev & test \\ \midrule

          \textbf{Offline} &  &  & \\
          \ UniLSTM - Global AED & \multirow{6}{*}{--} & 12.9 & 11.9 \\
          \ BLSTM - Global AED ({\tt T1}) & & \phantom{0}9.7 & \phantom{0}9.5 \\
          \ BLSTM - RNN-T ({\tt T2}) & & \phantom{0}9.4 & \phantom{0}9.2 \\
          \ BLSTM - MoChA & & 13.8 & 12.6 \\
          \ \ + QR ({\tt T3}) & & \bf{10.8} & \bf{\phantom{0}9.8} \\
          \ \ + CTC-ST & & 11.0 & 10.2 \\ \midrule

          \textbf{Streaming (UniLSTM encoder)} &  &  & \\
          \ RNN-T ({\tt T4}) & -- & 12.8 & 13.0 \\
          \ MoChA + QR ({\tt T5}) & -- & 16.7 & 15.0 \\
          \ MoChA + CTC-ST ({\tt T6}) & {\tt T5} & \bf{14.8} & \bf{13.2} \\
          \cdashlinelr{1-4}

          \textbf{Streaming (LC-BLSTM-40+20 encoder)} &  &  & \\
          \ RNN-T & {\tt T2} & 10.5 & 10.4 \\
          \ MoChA + QR & {\tt T3} & 13.8 & 12.2 \\
          \ MoChA + CTC-ST & {\tt T3} & \bf{12.0} & \bf{10.5} \\
          \cdashlinelr{1-4}

          \textbf{Streaming (LC-BLSTM-40+40 encoder)} &  &  & \\
          \ RNN-T & {\tt T2} & 10.0 & 10.0 \\
          \ MoChA + QR ({\tt T7}) & {\tt T3} & 13.4 & 11.3 \\
          \ MoChA + CTC-ST ({\tt T8}) & {\tt T3} & \bf{11.5} & \bf{\phantom{0}9.9} \\
          \bottomrule
    \end{tabular}
    }
    \endgroup
\end{table}

\begin{figure}[t]
  \centering
  \includegraphics[width=0.99\linewidth]{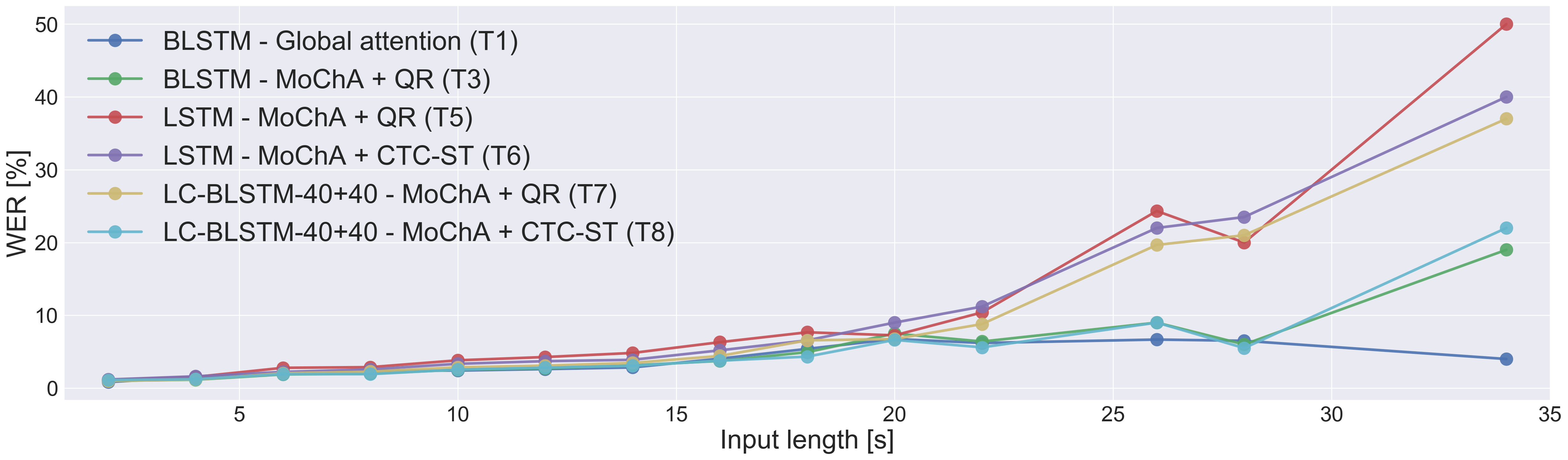}
  \caption{WER bucketed by different input length on the TEDLIUM2 \textit{test} set}
  \label{fig:taslp2021_wer_dist}
\end{figure}

\section{Results}\label{sec:taslp2021_results}
\subsection{TEDLIUM2}\label{ssec:result_tedlium2}
\subsubsection{Effectiveness of CTC-ST}
Table~\ref{tab:taslp2021_result_tedlium2} summarizes the results on the TEDLIUM2 corpus.
In the offline scenario, the naive implementation of MoChA showed very poor performance.
Quantity regularization (QR) drastically improved the performance ({\tt T3}).
Although CTC-ST also improved the performance, it was less effective than QR when it was applied from scratch.
The BLSTM RNN-T outperformed the global AED model.

In the streaming scenario, however, our proposed CTC-ST significantly improved the baseline performance as compared to QR regardless of the encoder type.
We obtained relative word error rate (WER) improvements of 12.0, 13.9, and 12.3\% for the UniLSTM, LC-BLSTM-40+20, and LC-BLSTM-40+40 MoChA models, respectively, on the \textit{test} set.
Although using a larger number of future lookahead frames improved the WER as expected, the effectiveness of CTC-ST was consistent.
The token boundaries from the MoChA and CTC branches in {\tt T5} and {\tt T6} are visualized in Fig.~\ref{fig:taslp2021_attention_plot}.
We can see that the MoChA boundaries moved to the left, and that the gap in the timing to emit tokens in both branches was reduced.
This is desirable for reducing the perceived latency~\cite{inaguma2020streaming}, which will be evaluated in Section~\ref{ssec:latency_ctc_sync}.
Moreover, we found that the CTC spikes slightly shifted to the left.
Although RNN-T was more robust for long-form utterances on the \textit{dev} set, MoChA optimized with CTC-ST matched the performance of RNN-T on the \textit{test} set.

\begin{table}[t]
    \centering
    \begingroup
    \caption{Results of the regularization methods and curriculum learning strategy on \underline{TEDLIUM2} with the LC-BLSTM-40+40 encoder configuration}\label{tab:taslp2021_result_tedlium2_curriculum}
    \scalebox{1.0}{
    \begin{tabular}{@{}cccccc@{}} \toprule
      \multirow{2}{*}{Initialization} & \multirow{2}{*}{Seed} &  \multirow{2}{*}{$\lossqua$} & \multirow{2}{*}{$\losssync$} & \multicolumn{2}{c}{WER [$\%$] ($\downarrow$)} \\ \cmidrule(lr){5-6}
      & & & & dev & test \\ \midrule

      \multirow{2}{*}{\shortstack{From scratch}} & \multirow{2}{*}{--} & \cmark & \xmark & 13.8 & 12.3 \\
      & & \xmark & \cmark & \bf{12.6} & \bf{10.9} \\
      \cmidrule(lr){1-6}

      \multirow{4}{*}{\shortstack{Curriculum\\learning}} & \multirow{4}{*}{{\tt T3}} & \xmark & \xmark & 18.1 & 16.9 \\
      & & \cmark & \xmark & 13.4 & 11.3 \\
      & & \xmark & \cmark & \bf{11.5} & \bf{\phantom{0}9.9} \\
      & & \cmark & \cmark & \bf{11.5} & 10.1 \\
      \bottomrule
    \end{tabular}
    }
    \endgroup
\end{table}

\vspace{2mm}
\subsubsection{Bucketing by input length}
Figure~\ref{fig:taslp2021_wer_dist} shows a plot of the WER bucketed by the input length on the \textit{test} set.
The plot confirms that the largest gains by CTC-ST were for utterances longer than 20 seconds.
The offline global AED model with joint CTC decoding ({\tt T1}) did not have difficulty in recognizing long utterances, whereas the baseline streaming MoChA models did, regardless of the encoder type ({\tt T5}, {\tt T7}).
As we had expected, the proposed CTC-ST successfully mitigated this problem ({\tt T6}, {\tt T8}).

\vspace{2mm}
\subsubsection{Effectiveness of curriculum learning}
Next, we investigate the effectiveness of the regularization methods (CTC-ST and QR) and the curriculum learning strategy by using LC-BLSTM-40+40 MoChA.
Table~\ref{tab:taslp2021_result_tedlium2_curriculum} summarizes the results.
Initialization with the offline model {\tt T3} was very helpful for both regularization methods, which is consistent with a previous study~\cite{adaptive_mocha}.
However, regularization with the CTC-ST loss or quantity loss in stage 2 was essential for achieving a performance gain even with curriculum learning.
When using the LC-BLSTM encoder, CTC-ST was more effective than QR regardless of the use of curriculum learning, unlike the results listed in Table~\ref{tab:taslp2021_result_tedlium2} for the BLSTM encoder.
Combining both losses did not lead to any further improvement, although it was more effective than applying the quantity loss alone.
Therefore, CTC-ST has an overlapping effect of encouraging MoChA to learn the scale of $\alphaij$ properly.

\begin{table}[t]
    \centering
    \begingroup
    \caption{Results with SpecAugment on \underline{TEDLIUM2}}\label{tab:taslp2021_result_tedlium2_specaugment}
    \scalebox{1.0}{
    \begin{tabular}{@{}lccc@{}} \toprule
     \multirow{2}{*}{Model} & \multirow{2}{*}{Seed} & \multicolumn{2}{c}{WER [$\%$] ($\downarrow$)} \\ \cmidrule(lr){3-4}
     & & dev & test \\ \midrule

       \textbf{Offline} & \\
       \ Transformer~\cite{karita2019comparative} & -- & \phantom{0}9.3 & \phantom{0}8.1 \\
       \ Transformer + semantic mask~\cite{semantic_mask} & -- & \phantom{0}-- & \phantom{0}7.7 \\
       \ BLSTM - Global AED~\cite{zeyer2019comparison} & -- & 10.3 & \phantom{0}8.8 \\
       \ BLSTM - Global AED & -- & \phantom{0}8.1 & \phantom{0}7.5 \\
       \ BLSTM - RNN-T & -- & \phantom{0}8.5 & \phantom{0}7.8 \\
       \midrule

       \textbf{Streaming (UniLSTM)} & \\
       \ RNN-T & {\tt T4} & \bf{11.6} & \bf{11.7} \\
       \ MoChA + QR & {\tt T5} & 17.7 & 16.0 \\
       \ MoChA + CTC-ST & {\tt T5} & \bf{12.9} & \bf{11.5} \\  %
       \cdashlinelr{1-4}

       \textbf{Streaming (LC-BLSTM-40+40)} &  \\
       \ RNN-T & {\tt T2} & \bf{\phantom{0}8.9} & \bf{\phantom{0}8.5} \\
       \ MoChA + QR & {\tt T3} & 12.9 & 11.2 \\
       \ MoChA + CTC-ST & {\tt T3} & \bf{10.6} & \bf{\phantom{0}8.6} \\
       \bottomrule
    \end{tabular}
    }
    \endgroup
\end{table}

\vspace{2mm}
\subsubsection{Combination with SpecAugment}
We next investigate the combination of CTC-ST and SpecAugment, whose results are summarized in Table~\ref{tab:taslp2021_result_tedlium2_specaugment}.
We set ($\specaugfreq$, $\specaugtime$) to (13, 50) for UniLSTM MoChA, (27, 50) for LC-BLSTM MoChA, and (27, 100) for the other models.
We used the same configurations on other corpora as well.
Because of the convergence issue, we applied SpecAugment only in stage 2 only.
\textcolor{\colorblue}{
Moreover, we increased $\lambdasync$ to 4.0 for the UniLSTM MoChA model.}\footnote{We found that this was beneficial when using SpecAugment for a weak encoder like UniLSTM.}
The naive streaming MoChA models only with QR did not obtain any improvement with SpecAugment, whereas the performance of the RNN-T models improved.
Therefore, this was a problem on the decoder side, rather than the encoder side.
However, CTC-ST mitigated this problem and showed additional 12.8\% and 13.1\% relative improvements for the UniSLTM and LC-BLSTM models, respectively, on the \textit{test} set.
Finally, MoChA optimized with CTC-ST matched the performance of RNN-T on the \textit{test} set when SpecAugment was applied as well.

\begin{table}[t]
    \centering
    \begingroup
    \caption{Results on \underline{Librispeech}. SpecAugment was used for all models. QR: quantity regularization. CTC-ST: CTC-synchronous training.}\label{tab:taslp2021_result_librispeech}
    \scalebox{0.98}{
    \begin{tabular}{@{}lcccc@{}} \toprule
      \multirow{3}{*}{Model} & \multicolumn{4}{c}{WER [$\%$] ($\downarrow$)} \\
      \cmidrule(lr){2-5}
       & \multicolumn{2}{c}{dev} & \multicolumn{2}{c}{test} \\
        & clean & other & clean & other \\ \midrule

        \textbf{Offline} & \\
        \ BLSTM - Global AED & 2.5 & \phantom{0}7.2 & 2.6 & \phantom{0}7.5 \\
        \ BLSTM - RNN-T & 2.8 & \phantom{0}8.1 & 3.1 & \phantom{0}8.7 \\
        \midrule

        \textbf{Streaming (UniLSTM)} & \\
        \ PTDLSTM - Triggered attention~\cite{moritz2019streaming_asru2019} & 5.6 & 16.2 & 5.7 & 16.9 \\
        \ MoChA + MWER~\cite{kim2020attention} & -- & \phantom{0}-- & 5.6 & 15.5 \\
        \ MoChA + \{char, BPE\}-CTC~\cite{garg2019improved} & -- & \phantom{0}-- & 4.4 & 15.2 \\
        \ RNN-T & \bf{3.7} & 11.6 & \bf{4.0} & 11.6 \\
        \ MoChA + QR & 4.5 & 11.4 & 4.7 & 11.9 \\  %
        \ MoChA + CTC-ST & \bf{3.8} & \bf{10.9} & \bf{4.0} & \bf{11.2} \\  %
        \cdashlinelr{1-5}

        \textbf{Streaming (LC-BLSTM-64+32)} & \\
        \ sMoChA~\cite{online_hybrid_ctc_attention} & -- & \phantom{0}-- & 6.0 & 16.7 \\
        \ MTA~\cite{online_hybrid_ctc_attention_taslp2020} & -- & \phantom{0}-- & 4.2 & 13.3 \\
        \cdashlinelr{1-5}

        \textbf{Streaming (LC-BLSTM-40+40)} &  \\
        \ RNN-T & \bf{3.3} & \phantom{0}9.7 & \bf{3.5} & 10.1 \\
        \ MoChA + QR & 3.9 & \phantom{0}9.0 & 4.8 & \phantom{0}9.3 \\
        \ MoChA + CTC-ST & \bf{3.3} & \bf{\phantom{0}8.8} & \bf{3.5} & \bf{\phantom{0}9.1} \\
        \bottomrule
    \end{tabular}
    }
    \endgroup
\end{table}

\subsection{Librispeech}
Table~\ref{tab:taslp2021_result_librispeech} summarizes the results on the LibriSpeech corpus.
For the UniLSTM MoChA model, CTC-ST achieved relative improvements of 14.8\% and 5.8\% on the \textit{test-clean} and \textit{test-other} sets, respectively.
For the LC-BLSTM MoChA model, we obtained gains of 27.0\% and 2.1\% with CTC-ST on the \textit{test-clean} and \textit{test-other} sets, respectively.
Therefore, we conclude that CTC-ST is effective for large-scale data as well.
Furthermore, the best MoChA models outperformed their RNN-T counterparts except for the \textit{dev-clean} set.

We also compared our models with streaming RNN-based E2E systems reported in the literature.
Our enhanced MoChA models optimized with CTC-ST showed the best performance.
To compare our model with sMoChA\cite{online_hybrid_ctc_attention} and MTA~\cite{online_hybrid_ctc_attention_taslp2020}, we deactivated SpecAugment in the LC-BLSTM MoChA model.
The results were 3.9\% and 11.2\% on the \textit{test-clean} and \textit{test-other} sets, and we still confirmed that our method outperformed them.
The average lookahead latency of our LC-BLSTM MoChA was 660ms (= 400ms ($\blocksizecurrent$)/2 + 400ms ($\blocksizeright$) + 60ms (CNN) while that was  640ms (= 640ms/2 + 320ms) in~\cite{online_hybrid_ctc_attention,online_hybrid_ctc_attention_taslp2020}.
Therefore, we consider that the difference in the lookahead latency is negligible.
Because CTC-ST is a method to enhance attention-based decoders, any type of encoder can be used.
Further improvement is expected by adapting Transformer~\cite{vaswani2017attention} and Conformer~\cite{gulati2020} encoders, which we leave for a future work.

\begin{table}[t]
    \centering
    \begingroup
    \caption{Results on \underline{CSJ}. SpecAugment was used for all models.}\label{tab:taslp2021_result_csj}
    \scalebox{1.0}{
    \begin{tabular}{@{}llccc@{}} \toprule
     \multirow{2}{*}{Model} & \multicolumn{3}{c}{WER / CER [$\%$] ($\downarrow$)} \\ \cmidrule(lr){2-4}
     & eval1 & eval2 & eval3 \\ \midrule

          \textbf{Offline} & \\
          \ \shortstack{CNN-TDNN-LSTM + \\LF-MMI$\to$sMBR~\cite{kanda2018lattice}} & \phantom{0}7.5 / 5.7 & 6.2 / 4.9 & 6.2 / 4.5 \\
          \ BLSTM - Global AED & \phantom{0}6.9 / 5.5 & 5.3 / 4.2 & 5.9 / 4.7 \\
          \midrule

          \textbf{Streaming (UniLSTM)} & \\
          \ MoChA + QR & \phantom{0}9.1 / 7.1 & 7.0 / {\bf 5.1} & 7.6 / 5.8 \\
          \ MoChA + CTC-ST & \phantom{0}{\bf 8.7} / {\bf 6.4} & {\bf 6.4} / 5.3 & {\bf 7.2} / {\bf 5.6} \\
          \cdashlinelr{1-4}

          \textbf{Streaming (LC-BLSTM-40+40)} & \\
          \ MoChA + QR & \phantom{0}7.6 / 6.0 & 5.9 / 4.8 & 6.5 / 5.0 \\
          \ MoChA + CTC-ST & \phantom{0}{\bf 7.4} / {\bf 5.8} & {\bf 5.6} / {\bf 4.5} & {\bf 6.4} / {\bf 4.9} \\
          \bottomrule
    \end{tabular}
    }
    \endgroup
\end{table}

\subsection{CSJ}
Table~\ref{tab:taslp2021_result_csj} summarizes the results on the CSJ.
For both UniLSTM and LC-BLSTM MoChA models, we observed clear improvements with CTC-ST, which was consistent with the previous experiments.
However, the relative gains for LC-BLSTM MoChA were smaller than those in other corpora.
We reason that this was because the utterance lengths in the CSJ are relatively shorter, as shown in Fig.~\ref{fig:taslp2021_corpus_utt_dist}.
We will analyze this behaviour by simulating long-form speech utterances in Section~\ref{sec:longform_evaluation}.

\begin{table}[t]
    \centering
    \begingroup
    \caption{Results on \underline{AMI (SDM)}. SpecAugment was used for all models.}\label{tab:taslp2021_result_ami}
    \scalebox{1.0}{
    \begin{tabular}{@{}lcc@{}} \toprule
      \multirow{2}{*}{Model} & \multicolumn{2}{c}{WER [$\%$] ($\downarrow$)} \\ \cmidrule(lr){2-3}
       & dev & eval \\ \midrule
          \textbf{Offline} & \\
          \ TDNN + LF-MMI~\cite{peddinti2016far} & 42.8 & 46.6 \\
          \ BLSTM - Global AED & \bf{37.4} & \bf{40.1} \\
          \ BLSTM - RNN-T & 38.1 & 41.2 \\
          \cdashlinelr{1-3}

          \textbf{Streaming (LC-BLSTM-40+40)} & \\
          \ RNN-T & 41.1 & 44.1 \\
          \ MoChA + QR & 46.5 & 49.7 \\
          \ MoChA + CTC-ST & \bf{40.6} & \bf{43.0} \\ %
          \bottomrule
    \end{tabular}
    }
    \endgroup
\end{table}

\subsection{AMI}
Table~\ref{tab:taslp2021_result_ami} summarizes the results on the AMI corpus.
We did not use any external LMs on this corpus because we did not observe any improvement.
We observed significant improvement with CTC-ST in the far-field ASR task as well.
CTC-ST gave 12.6\% and 13.4\% relative improvements over the baseline LC-BLSTM MoChA model on the \textit{dev} and \textit{eval} sets, respectively.
LC-BLSTM MoChA also achieved better performance than that of LC-BLSTM RNN-T.
Note that our baseline was very strong, as demonstrated by its superior performance compared to the TDNN+LF-MMI system~\cite{peddinti2016far}.
As shown in Fig.~\ref{fig:taslp2021_corpus_utt_dist}, utterance lengths in the AMI corpus are relatively short compared to other corpora.
Therefore, we can conclude that CTC-ST is also very effective for noisy speech, for which AED models have trouble learning alignments.

\section{\textcolor{\colorblue}{Evaluation of robustness to long-form speech}}\label{sec:longform_evaluation}
In Section~\ref{ssec:result_tedlium2}, we have observed that CTC-ST was effective for reducing WER of long-form utterances on TEDLIUM2.
In this section, we further analyze this behavior by simulating long-form evaluation sets on other domains.
We used CSJ and Librispeech for this purpose because input lengths of the original utterances in the evaluation sets were seen during training, as shown in Fig.~\ref{fig:taslp2021_corpus_utt_dist}.
We simulated long-form utterances by merging adjacent utterances according to timestamps.
Specifically, given a maximum input length threshold $\tcat$ [sec.], we concatenated adjacent utterances of the same speaker from the first utterance in a greedy way until the accumulated utterance length surpassed $\tcat$.
We continued this process until no segment was merged in an iteration.

\begin{figure}[t]
  \centering
  \includegraphics[width=0.99\linewidth]{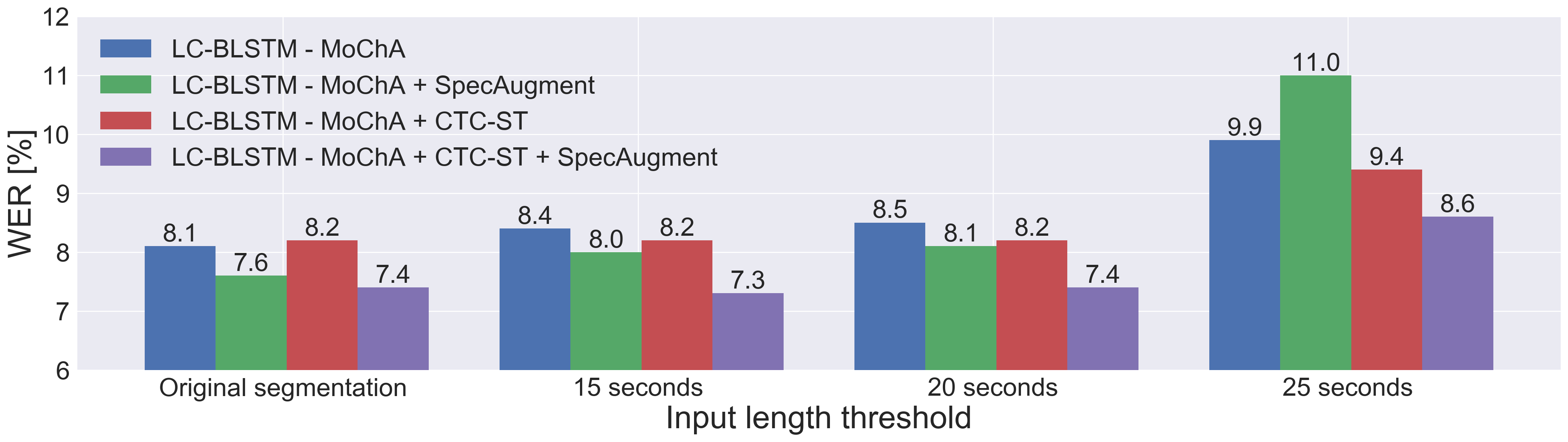}
   \caption{WER of simulated long-form speech on the CSJ \textit{eval1} set, with the different maximum input length threshold $\tcat$. LC-BLSTM-40+40 encoder was used.}
  \label{fig:taslp2021_wer_csj_input_length}
\end{figure}

\begin{figure}[t]
  \centering
  \includegraphics[width=0.99\linewidth]{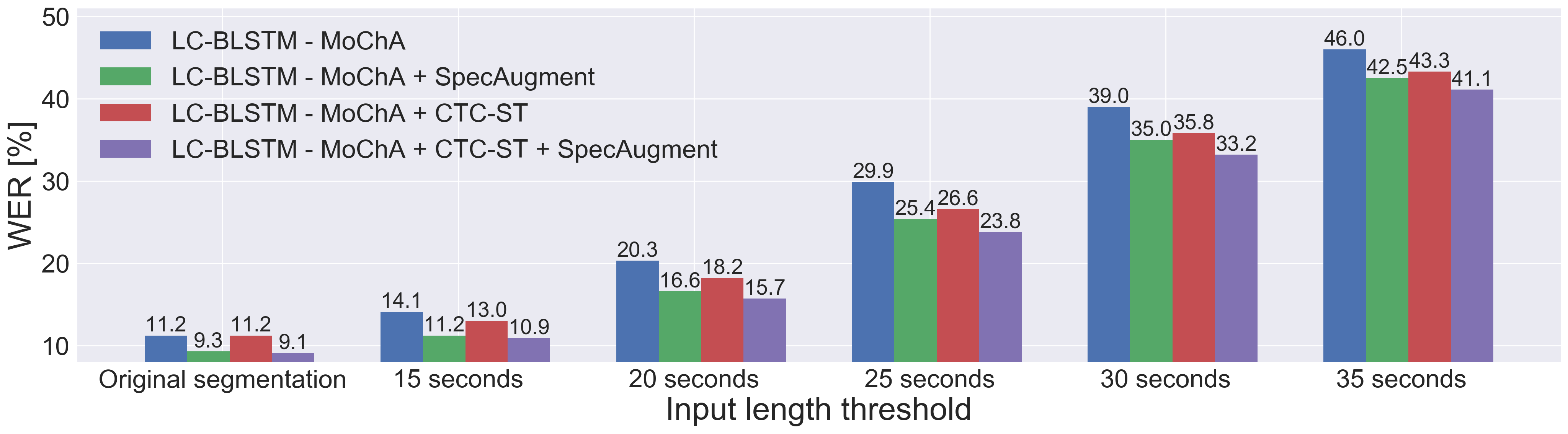}
   \caption{\textcolor{\colorblue}{WER of simulated long-form speech on the Librispeech \textit{test-other} set, with the different maximum input length threshold $\tcat$. LC-BLSTM-40+40 encoder was used.}}
  \label{fig:taslp2021_wer_libri_input_length}
\end{figure}

Figures~\ref{fig:taslp2021_wer_csj_input_length} and ~\ref{fig:taslp2021_wer_libri_input_length} show the results on CSJ and Librispeech, respectively.
On CSJ, the baseline MoChA model without SpecAugment (blue bars) performed well with manual audio segmentation.
However, as $\tcat$ increased, the performance was gradually degraded, whereas MoChA models trained with CTC-ST only (red bars) were robust in recognizing the long-form utterances.
On the other hand, the WER of the naive MoChA trained with SpecAugment (green bars) was increased quickly for longer utterances.
This indicates that SpecAugment affected the training of the naive MoChA model, which could not be observed with the original test sets because they did not include unseen input lengths.
However, we confirmed that CTC-ST mitigated this problem, showing the lowest WER on all lengths (purple bars).
We also confirmed the effectiveness of CTC-ST on Librispeech in all length bins as well although SpecAugment without CTC-ST did not degrade WER as severely as on CSJ.
Still, the gains by CTC-ST were larger in long-form speech.
Therefore, we can conclude that CTC-ST is effective for recognizing long-form speech, which is generally challenging for AED models~\cite{chiu2019comparison,hsiao2020online}.

\section{Evaluation of emission latency}\label{sec:emission_latency_evaluation}
In this section, we evaluate the emission latency and compare CTC-ST with alignment knowledge distillation from a hybrid system~\cite{inaguma2020streaming}.
Moreover, we also compare MoChA with RNN-T.

\subsection{Emission latency metric}\label{ssec:latency_metric}
\subsubsection{Token emission latency (TEL)}
Unlike the algorithmic latency introduced by lookahead frames in the encoder, the token emission latency (TEL) represents the user-perceived latency in a real application~\cite{shangguan2021dissecting}.
Although some previous works investigated the endpoint latency corresponding the last token in voice search and assistance tasks~\cite{li2020towards,sainath2020streaming,shangguan2021dissecting}, we mainly focus here on the \textit{per-token} emission latency because we are interested in long-form speech applications, as in lectures and meetings.

Following~\cite{inaguma2020streaming}, we define the TEL as the difference in timing between the reference and predicted boundaries.
To obtain the reference token boundaries, we perform forced alignment with Kaldi.
The predicted boundaries in an utterance are obtained from the input timesteps at which monotonic attention in MoChA is activated, {\ie}, $\{j|\zij=1\}_{i=1,\ldots,U}$.
The TEL of the $i$-th token in the $n$-th utterance, $\Delta_{n,i}$, is calculated as
\begin{eqnarray}
\Delta_{n,i} [ms]=\hat{\rm{b}}_{i}^{n} - {\rm b}_{i}^{n},
\end{eqnarray}
where $\hat{\rm{b}}_{i}^{n}$ and ${\rm b}_{i}^{n}$ are the $i$-th predicted and reference boundaries, respectively, in the $n$-th utterance.
We do not include the $\langle$eos$\rangle$ token for TEL calculation.
A negative latency can be observed for some tokens because of premature boundary detection.
To match the lengths of a hypothesis and the corresponding reference when calculating the TEL, we apply teacher forcing by conditioning the decoder on the ground-truth transcript.
However, the WER is reported with beam search decoding using the same model.
Therefore, the TEL is a \textit{corpus-level} latency metric.
In the following, we report the median (\textit{PT@50}) and 90th (\textit{PT@90}) percentile values of the corpus-level TEL distributions.

We also evaluate the TEL of CTC.
In this case, we perform forced alignment and use the most plausible alignment path to calculate the TEL.

\begin{table}[t]
    \centering
    \begingroup
    \caption{Token emission latency (TEL) on \underline{TEDLIUM2}, with the columns $\alignment$ indicating the alignment. UniLSTM encoder was used for all models. TEL was calculated on the \textit{test} set. PT@X represents the X-th percentile. $\deltadecot$ corresponds to the number of encoder outputs [40ms]. ${}^{\diamondsuit}$SpecAugment was not used.}\label{tab:taslp2021_result_emission_latency_tedlium2}
    \scalebox{0.98}{
    \begin{tabular}{@{}clcccc@{}} \toprule
     \multirow{2}{*}{$\alignment$} & \multirow{2}{*}{Model} & \multicolumn{2}{c}{WER [$\%$] ($\downarrow$)} & \multicolumn{2}{c}{TEL [ms] ($\downarrow$)} \\ \cmidrule(lr){3-4} \cmidrule(lr){5-6}
     & & dev & test & PT@50 & PT@90 \\ \midrule

      \multirow{4}{*}{--} & MoChA + QR${}^{\diamondsuit}$ & 16.7 & 15.0 & 280 & 680 \\
      & \ + from CTC branch & 15.1 & 14.7 & 200 & 280 \\  %

      & MoChA + QR & 17.7 & 16.0 & 320 & 840 \\  %
      & \ + from CTC branch & 13.3 & 12.8 & 240 & 360 \\  %
      \midrule

     \multirow{15}{*}{CTC} & MoChA + CTC-ST & \\
      & \ - $\lambdasync=0.5$ & 13.0 & 11.6 & 240 & 400 \\ %
      & \ \ \ + from CTC branch & 13.4 & 12.9 & 200 & 320 \\ %
      \cdashlinelr{2-6}

      & \ - $\lambdasync=1.0$ & 13.6 & 11.6 & 200 & 360 \\
      & \ \ \ + from CTC branch & 13.6 & 13.1 & 200 & 280 \\  %
    \cdashlinelr{2-6}

      & \ - $\lambdasync=2.0$ & 13.3 & 11.5 & \bf{120} & 280 \\
      & \ \ \ + from CTC branch & 13.7 & 13.2 & 120 & 240 \\  %
      \cdashlinelr{2-6}

      & \ - $\lambdasync=3.0$ & 13.0 &\bf{11.4} & \bf{120} & \bf{240} \\
      & \ \ \ + from CTC branch & 14.0 & 13.4 & 120 & 240 \\  %
      \cdashlinelr{2-6}

      & \ - $\lambdasync=4.0$ & \bf{12.9} & 11.5 & \phantom{0}\bf{80} & \bf{240} \\
      & \ \ \ + from CTC branch & 14.1 & 13.8 & 120 & 200 \\  %
      \cdashlinelr{2-6}

      & \ - $\lambdasync=5.0$ & 13.4 & 11.9 & \phantom{0}\bf{80} & \bf{200} \\
      & \ \ \ + from CTC branch & 14.6 & 14.5 & \phantom{0}80 & 200 \\  %
     \midrule

     \multirow{10}{*}{\shortstack{Hybrid\\ASR}}
     & MoChA + DeCoT & \\
     & \ - $\deltadecot=8$ & 13.4 & 11.5 & \bf{160} & \bf{240} \\
     & \ - $\deltadecot=12$ & 12.9 & 11.2 & 200 & 320 \\
     & \ - $\deltadecot=16$ & \bf{12.5} & \bf{11.0} & 280 & 440 \\
     & \ - $\deltadecot=20$ & 13.0 & 11.3 & 240 & 400 \\
     & \ - $\deltadecot=24$ & 13.5 & 11.7 & 280 & 480 \\
    \cdashlinelr{2-6}

     & MoChA + MinLT & \\
     & \ - $\lambdaminlt=1.0$ & 14.4 & 12.7 & 240 & 520 \\
     & \ - $\lambdaminlt=2.0$ & 13.5 & 11.7 & 240 & 360 \\
     & \ - $\lambdaminlt=3.0$ & 13.9 & 11.6 & 200 & 320 \\
     \bottomrule
    \end{tabular}
    }
    \endgroup
\end{table}

\begin{table}[t]
    \centering
    \begingroup
    \caption{Token emission latency (TEL) on \underline{Librispeech}. UniLSTM encoder was used for all models. TEL was averaged over the \textit{test-clean} and \textit{test-other} sets. ${}^{\diamondsuit}$SpecAugment was not used.}\label{tab:taslp2021_result_emission_latency_librispeech}
    \scalebox{0.98}{
    \begin{tabular}{@{}clcccc@{}} \toprule
     \multirow{3}{*}{$\alignment$} & \multirow{3}{*}{Model} & \multicolumn{2}{c}{WER [$\%$] ($\downarrow$)} & \multicolumn{2}{c}{TEL [ms] ($\downarrow$)} \\ \cmidrule(lr){3-4} \cmidrule(lr){5-6}
     & & \multicolumn{2}{c}{test} & \multirow{2}{*}{PT@50} & \multirow{2}{*}{PT@90} \\
     & & clean & other &  &  \\
     \midrule

     \multirow{4}{*}{--} & MoChA + QR${}^{\diamondsuit}$ & 4.8 & 14.2 & 320 & 520 \\  %
     & \ + from CTC branch & 6.2 & 17.5 & 280 & 400 \\  %

     & MoChA + QR & 4.7 & 11.9 & 360 & 560 \\  %
     & \ + from CTC branch &5.3 & 14.4 & 320 & 440 \\  %
     \midrule

     \multirow{8}{*}{CTC} & MoChA + CTC-ST \\
     & \ - $\lambdasync=1.0$ & \bf{4.0} & \bf{11.2} & 240 & 400 \\  %
     & \ \ \ + from CTC branch & 5.3 & 14.6 & 240 & 360 \\  %
     \cdashlinelr{2-6}

     & \ - $\lambdasync=2.0$ & \bf{4.0} & 11.5 & \bf{160} & \bf{320} \\
     & \ \ \ + from CTC branch & 5.7 & 15.2 & 160 & 320 \\  %
     \cdashlinelr{2-6}

     & \ - $\lambdasync=3.0$ & 4.1 & 11.7 & \bf{120} & \bf{280} \\  %
     & \ \ \ + from CTC branch & 6.0 & 15.7 & 120 & 280 \\  %
     \midrule

     \multirow{7}{*}{\shortstack{Hybrid\\ASR}}
     & MoChA + DeCoT \\
    & \ - $\deltadecot=12$ &3.9 & 11.6 & \bf{240} & \bf{360} \\  %
     & \ - $\deltadecot=16$ & 3.9 & 11.4 & 320 & 440 \\  %
    & \ - $\deltadecot=20$ &4.0 & 11.2 & 360 & 480 \\  %
     \cdashlinelr{2-6}

     & MoChA + MinLT \\
     & \ - $\lambdaminlt=1.0$ & 4.5 & 11.7 & 320 & 400 \\  %
     & \ - $\lambdaminlt=2.0$ & 5.6 & 12.4 & 280 & 400 \\  %
     \bottomrule
    \end{tabular}
    }
    \endgroup
\end{table}

\vspace{2mm}
\subsubsection{Word emission latency (WEL)}
\textcolor{\colorblue}{
To enable a comparison of the emission latency between MoChA and RNN-T, which have different output units in our experiments, we also evaluate the word emission latency (WEL).
The WEL only considers the time differences of the last subword in each word.
When calculating the WEL of RNN-T, we conduct forced alignment similarly to CTC and use the most plausible alignment path.
Moreover, we also evaluate the three kinds of WEL; (1) per-word WEL (\textit{average WEL}), (2) WEL corresponding to the first word in an utterance (\textit{first WEL})~\cite{shangguan2021dissecting}, and (3) WEL corresponding to the last word (\textit{last WEL})~\cite{yu2021fastemit,yu2021dualmode}.
}

\subsection{Latency evaluation of CTC-ST}\label{ssec:latency_ctc_sync}
As shown in Fig.~\ref{fig:taslp2021_attention_plot}, the naive MoChA tended to emit tokens later than the corresponding CTC spikes, and CTC-ST reduced the gap in the example.
To evaluate this quantitatively, we calculated the TEL.
We used UniLSTM encoders for this purpose, because LC-BLSTM encoders introduce the algorithmic latency on the encoder side, whereas we are interested in the emission latency on the decoder side.
Tables~\ref{tab:taslp2021_result_emission_latency_tedlium2} and~\ref{tab:taslp2021_result_emission_latency_librispeech} summarize the results on the TEDLIUM2 and Librispeech corpora, respectively.
The TEL on Librispeech was averaged over the \textit{test-clean} and \textit{test-other} sets.
\textcolor{\colorblue}{
We first observed that SpecAugment significantly increased the TEL of the baseline MoChA model.
We also evaluated the TEL from the CTC branch\footnote{The corresponding WER was calculated with a beam width of 10 and shallow fusion.}, and it also increased slightly by applying SpecAugment.
On the other hand, we confirmed that CTC-ST significantly reduced both TEL and WER on both corpora.
Increasing $\lambdasync$ up to 4.0 showed improvements of both metrics on TEDLIUM2.
On the other hand, on Librispeech, WER was best at $\lambdasync=1.0$ while the TEL was continuously reduced by increasing $\lambdasync$ up to 3.0.
PT@50 of the baseline MoChA with SpecAugment was reduced by 240ms and 240ms on TEDLIUM2 and Librispeech, respectively.
PT@90 was reduced by 600ms and 280ms on TEDLIUM2 and Librispeech, respectively.
The TEL of MoChA matched that of CTC in most conditions, confirming the function of CTC-ST.
CTC-ST traded WER and TEL effectively by changing $\lambdasync$.
Interestingly, we found that CTC-ST also reduced the TEL of the CTC branch by increasing $\lambdasync$.
This indicates that joint training reduced the TEL of the other branch interactively via the shared encoder.
Although CTC itself had a delay from the reference acoustic boundaries, it provided better timing to emit tokens for MoChA.
}

\subsection{Comparison with alignment distillation from hybrid system}\label{ssec:latency_hybrid}  %
Next, we compared CTC-ST with methods of alignment knowledge distillation methods from a hybrid system.
We trained MoChA models with MinLT and DeCoT on the same model configuration.
Tables~\ref{tab:taslp2021_result_emission_latency_tedlium2} and~\ref{tab:taslp2021_result_emission_latency_librispeech} summarize the results on the TEDLIUM2 and Librispeech corpora, respectively.
We observed that both DeCoT and MinLT also reduced the WER and TEL from those of the baseline model.\footnote{Unlike in ~\cite{inaguma2020streaming}, we applied SpecAugment to DeCoT and MinLT. We found that those methods can also tolerate noisy inputs to some extent.}
DeCoT with the optimal $\deltadecot$ outperformed MinLT in both metrics on both corpora.
CTC-ST also outperformed MinLT in both metrics.
On Librispeech, increasing $\lambdasync$ brought a large TEL reduction without hurting the WER so much while MinLT sacrificed the WER a lot with a small TEL reduction.
\textcolor{\colorblue}{
Compared to DeCoT, CTC-ST achieved a lower TEL, especially for PT@50,  with a comparable WER.
This was because DeCoT focused on tokens whose emission latency surpassed $\deltadecot$ and thus the TEL reduction was large in tail parts (PT@90).
On the other hand, CTC-ST reduced emission latency of all tokens.}
Therefore, we conclude that CTC-ST can achieve a similar or better tradeoff compared to alignment knowledge distillation from a hybrid system without relying on external alignment information.

\begin{table}[t]
    \centering
    \begingroup
    \caption{Word emission latency (WEL) of MoChA and RNN-T on \underline{TEDLIUM2}, with all models using SpecAugment}\label{tab:taslp2021_result_word_emission_latency_tedlium2}
    \scalebox{0.92}{
    \begin{tabular}{@{}lcccccc@{}} \toprule
     \multirow{2}{*}{Model} & \multicolumn{2}{c}{Avg WEL [ms] ($\downarrow$)} & \multicolumn{2}{c}{First WEL [ms]} & \multicolumn{2}{c}{Last WEL [ms]} \\ \cmidrule(lr){2-3} \cmidrule(lr){4-5} \cmidrule(lr){6-7}
     & PT@50 & PT@90 & PT@50 & PT@90 & PT@50 & PT@90 \\ \midrule

     RNN-T & 240 & 320 & 240 & 400 & \phantom{-0}0 & 160  \\
     MoChA & 320 & 840 & 320 & 640 & \phantom{-}80 & 440 \\
     \ + CTC-ST & \phantom{0}\bf{80} & \bf{240} & \phantom{0}\bf{80} & \bf{200} & \bf{-80} & \phantom{0}\bf{40} \\
     \bottomrule
    \end{tabular}
    }
    \endgroup
    \vspace{3mm}
\end{table}

\begin{table}[t]
    \centering
    \begingroup
    \caption{Word emission latency (WEL) of MoChA and RNN-T on \underline{Librispeech}, with all models using SpecAugment}\label{tab:taslp2021_result_word_emission_latency_librispeech}
    \scalebox{0.92}{
    \begin{tabular}{@{}lcccccc@{}} \toprule
     \multirow{2}{*}{Model} & \multicolumn{2}{c}{Avg WEL [ms] ($\downarrow$)} & \multicolumn{2}{c}{First WEL [ms]} & \multicolumn{2}{c}{Last WEL [ms]} \\ \cmidrule(lr){2-3} \cmidrule(lr){4-5} \cmidrule(lr){6-7}
     & PT@50 & PT@90 & PT@50 & PT@90 & PT@50 & PT@90 \\ \midrule

     RNN-T & 280 & 400 & 320 & 400 & 160 & 280 \\
     MoChA & 360 & 480 & 400 & 520 & 240 & 360 \\
     \ + CTC-ST & \bf{160} & \bf{280} & \bf{200} & \bf{320} & \phantom{00}\bf{0} & \bf{120} \\  %
     \bottomrule
    \end{tabular}
    }
    \endgroup
\end{table}

\subsection{Comparison with RNN-T}
\textcolor{\colorblue}{
Finally, we compare the emission latency between MoChA and RNN-T.
In addition to the average per-word statistics, we also calculated the WEL corresponding to the first and last tokens.
The results in Tables~\ref{tab:taslp2021_result_word_emission_latency_tedlium2} and~\ref{tab:taslp2021_result_word_emission_latency_librispeech} present the WEL on TEDLIUM2 and Librispeech, respectively.
We confirmed that MoChA trained with CTC-ST achieved lower WELs than those of RNN-T in all the conditions on both corpora.
Note that RNN-T was also jointly trained with the CTC objective, and thus can be regarded as a strong baseline.
Comparing the first WEL and the last WEL, we found that the latter had a lower latency.
We reason that more acoustic contexts were necessary to emit the first word because there was no linguistic context on the decoder side.
On the other hand, the last WEL of MoChA with CTC-ST was close to zero.
}

\begin{table}[t]
    \centering
    \begingroup
    \caption{Comparison of CTC alignment generation in CTC-ST on \underline{TEDLIUM2}, with all models using SpecAugment}\label{tab:taslp2021_result_precompute_tedlium2}
    \scalebox{1.1}{
    \begin{tabular}{@{}lcc@{}} \toprule
     \multirow{2}{*}{Model} & \multicolumn{2}{c}{WER [$\%$] ($\downarrow$)} \\ \cmidrule(lr){2-3}
     & dev & test \\ \midrule

     \textbf{UniLSTM - MoChA} & & \\
     \ on the fly & \bf{13.6} & \bf{11.6} \\
     \ precomputing & 13.9 & 12.0 \\
     \cmidrule(lr){1-3}

     \textbf{LC-BLSTM-40+40 - MoChA} & & \\
     \ on the fly & \bf{10.6} & \bf{\phantom{0}8.6} \\
     \ precomputing & 11.2 & \phantom{0}9.2 \\

     \bottomrule
    \end{tabular}
    }
    \endgroup
    \vspace{3mm}
\end{table}

\begin{table}[t]
    \centering
    \begingroup
    \caption{Comparison of CTC alignment generation in CTC-ST on \underline{Librispeech}, with all models using SpecAugment}\label{tab:taslp2021_result_precompute_Librispeech}
    \scalebox{1.1}{
    \begin{tabular}{@{}lcccc@{}} \toprule
     \multirow{3}{*}{Model} & \multicolumn{4}{c}{WER [$\%$] ($\downarrow$)} \\ \cmidrule(lr){2-5}
     & \multicolumn{2}{c}{dev} & \multicolumn{2}{c}{test} \\
     & clean & other & clean & other \\
     \midrule

     \textbf{UniLSTM - MoChA} & & \\
     \ on the fly & \bf{3.8} & 10.9 & \bf{4.0} & 11.2 \\
     \ precomputing & \bf{3.8} & \bf{10.7} & \bf{4.0} & \bf{11.1} \\
     \cmidrule(lr){1-5}

     \textbf{LC-BLSTM-40+40 - MoChA} & & \\
     \ on the fly & \bf{3.3} & \bf{\phantom{0}8.8} & \bf{3.5} & \bf{\phantom{0}9.1} \\
     \ precomputing & 3.4 & \bf{\phantom{0}8.8} & \bf{3.5} & \bf{\phantom{0}9.1} \\

     \bottomrule
    \end{tabular}
    }
    \endgroup
\end{table}

\section{Analysis}
In this section, we perform an ablation study of alignment generation in CTC-ST.
Finally, we compare MoChA and RNN-T in terms of the inference speed.

\subsection{Effect of incremental alignment update}\label{ssec:precompute_alignment}
In the above experiments, we generated the reference boundaries from CTC alignments with the model parameters at each training step \textit{on the fly}.
Here, we investigated the effect of using \textit{fixed} reference boundaries throughout stage 2 by using parameters optimized in stage 1.
We refer to this strategy as \textit{precomputing}.
When generating the CTC alignments for precomputing, we deactivated SpecAugment and other regularization methods such as dropout.
We used $\lambdasync = 1.0$ in this experiment.
Tables~\ref{tab:taslp2021_result_precompute_tedlium2} and \ref{tab:taslp2021_result_precompute_Librispeech} summarize the results on the TEDLIUM2 and Librispeech corpora, respectively.
For TEDLIUM2, the on-the-fly CTC alignment generation consistently outperformed the precomputing strategy, regardless of the encoder type.
Note, however, that the precomputing strategy also significantly outperformed the baseline listed in Table~\ref{tab:taslp2021_result_tedlium2_specaugment}.
For Librispeech, precomputing showed similar performances to those of on-the-fly computing.
This was because the parameters learned in stage 1 had already provided good CTC alignments by leveraging more training data.
This also confirms the observation that CTC-ST achieved similar WERs to those of DeCoT on Librispeech in Section~\ref{ssec:latency_hybrid}.

\begin{figure}[t]
  \centering
  \includegraphics[width=0.99\linewidth]{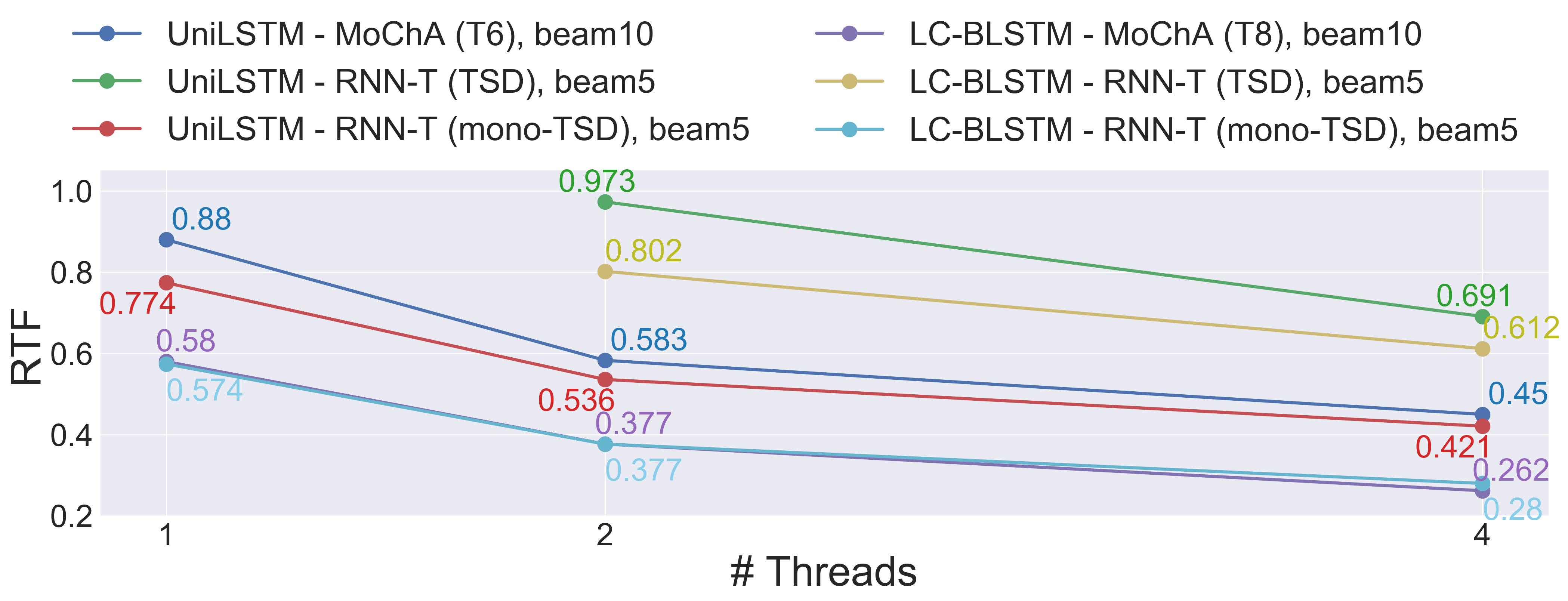}
  \vspace{-3mm}
  \caption{Inference speed as measured by the real time factor (RTF). TSD: time-synchronous decoding.}
  \label{fig:taslp2021_rtf}
\end{figure}

\subsection{Inference speed}\label{ssec:inference_speed}
While we have shown that MoChA can match RNN-T in terms of accuracy, we also evaluated its efficiency as compared to RNN-T in terms of the inference speed.
For both models, we precomputed token embeddings before decoding.
For RNN-T, we cached prediction network states corresponding to the same hypothesis~\cite{he2019streaming} and batched all hypotheses in the beam for updating the prediction network and joint network~\cite{saon2020alignment}.
We applied both TSD and mono-TSD as the search algorithm~\cite{tripathi2019monotonic}.
The maximum expansion number was set to 3 per frame in the TSD algorithm.
We used the best MoChA (optimized with CTC-ST) and RNN-T models trained with SpecAugment, with the beam width set to 10 and \{5, 10\}, respectively.
The inference speed was measured with a 6-core Intel(R) Xeon(R) Gold 6128 CPU @ 3.4GHz.
We investigated \{1, 2, 4\} threads, and we report the real-time factor (RTF) obtained by averaging five trials.
Figure~\ref{fig:taslp2021_rtf} shows the results on the TEDLIUM2 \textit{test} set.
We observed that all MoChA models achieved an RTF of less than 1.0 with a single thread in a Python implementation.
Using more threads led to faster decoding.
The UniLSTM encoder was slower than the LC-BLSTM encoder because of the incremental state update on a per-frame basis.
On the other hand, RNN-T required using the mono-TSD algorithm with the half beam width to achieve a similar inference speed.
Moreover, RNN-T with the TSD algorithm was much slower because of the multiple symbol expansions per frame.

\section{Conclusions}\label{sec:conclution}
In this article, we have proposed CTC synchronous training (CTC-ST), a self-distillation method for knowledge in input-output alignment to improve the performance of streaming AED models.
Specifically, we distill knowledge of token boundary positions from a CTC model to a MoChA model, both of which share an encoder and are trained jointly.
The proposed method forces MoChA to generate tokens in positions similar to those predicted by CTC, by synchronizing both sets of token boundaries during training.
Experimental evaluations on four benchmark datasets demonstrated that the proposed method significantly improved MoChA in terms of both the recognition accuracy and the emission latency, especially for long-form and noisy utterances.
We also compared the proposed method with methods of alignment knowledge distillation from an external hybrid ASR system and achieved a similar tradeoff of the accuracy and latency without any external alignments.
Finally, we showed that MoChA can achieve comparable recognition accuracy, lower emission latency, and faster inference speed compared to RNN-T.

In future work, we would like to further reduce the gap in recognition accuracy between RNN-T and MoChA in very long utterances.
Reducing flicker of MoChA by selecting stable partial hypotheses that do not change in the subsequent prefix expansion, which were studied for incremental systems~\cite{fugen2009system,mcgraw2012estimating,selfridge2011stability,nguyen2020low,liu2020,nguyen2020super} and RNN-T~\cite{shangguan2020}, is also an interesting research direction.

\bibliographystyle{IEEEtran}
\bibliography{reference}

\ifCLASSOPTIONcaptionsoff
  \newpage
\fi

\end{document}